\documentclass[pre]{revtex4-2}
\usepackage[utf8]{inputenc}
\usepackage{amsmath}
\usepackage{amsfonts}
\usepackage{graphicx}
\usepackage{subcaption}
\usepackage{physics}
\usepackage{hyperref}
\usepackage{xcolor}

\newcommand{\dr}[1]{{#1}}

\begin{document}

\title{A machine learning approach to fast thermal equilibration}
\author{Diego Rengifo}
\email{d.rengifo@uniandes.edu.co}
\affiliation{Departamento de Física, Universidad de los Andes,
  Bogotá, Colombia}
\affiliation{Institut für Physik und Astronomie, Technische Universität Berlin, Hardenbergstraße 36, D-10623 Berlin, Germany}
\author{Gabriel Téllez}
\email{gtellez@uniandes.edu.co}
\affiliation{Departamento de Física, Universidad de los Andes,
  Bogotá, Colombia}

\begin{abstract}
        We present a method to design driving protocols that achieve fast thermal equilibration of a system of interest using techniques inspired by machine learning training algorithms. For example, consider a Brownian particle manipulated by optical tweezers. The force on the particle can be controlled and adjusted over time, resulting in a driving protocol that transitions the particle from an initial state to a final state. Once the driving protocol has been completed, the system requires additional time to relax to thermal equilibrium. Designing driving protocols that bypass the relaxation period is of interest so that, at the end of the protocol, the system is either in thermal equilibrium or very close to it. Several studies have addressed this problem through reverse engineering methods, which involve prescribing a specific evolution for the probability density function of the system and then deducing the corresponding form of the driving protocol potential. Here, we propose a new method that can be applied to more complex systems where reverse engineering is not feasible. We simulate the evolution of a large ensemble of trajectories while tracking the gradients with respect to a parametrization of the driving protocol. The final probability density function is compared to the target equilibrium one. Using machine learning libraries, the gradients are computed via backpropagation and the protocol is iteratively adjusted until the optimal protocol is achieved. We demonstrate the effectiveness of our approach with several examples.
\end{abstract}

\maketitle

An important branch of physics and engineering is control theory~\cite{Bechhoefer_control21}, whose objective is to design protocols to control the evolution of a given system and drive it to the desired final state. Often, these protocols must also satisfy certain constraints, such as optimizing energy costs. In recent decades, technology has advanced to the point where we can manipulate microsized systems\dr{~\cite{WangEvans02, Levitodynamics2021, Trizac2023}}, such as colloidal particles with optical tweezers, micro-oscillators, and microscopic active matter. Due to their small size, significant thermal fluctuations affect these systems. As a result, effectively controlling such systems often requires oversight not only of the average state but also of the fluctuations. This has given rise to a research area where the statistical description of the system (e.g., the density matrix or the probability density function in phase space) becomes the central object of control.

Under time-dependent control, the probability distributions of these systems
evolve according to the Fokker-Planck equation. Once the control protocol is
completed, the system remains out of equilibrium and requires time to relax to
its new equilibrium state. In many cases, rapid control of the system is
desirable to bypass the relaxation period. Such protocols are known as
engineered swift equilibration (ESE)~\cite{trizac1} when targeting thermal
equilibrium, shortcuts to isothermality~\cite{geng17}, or generally as swift
state-to-state transformations. For a review,
see Ref.~\cite{guery-odelinDrivingRapidlyRemaining2023}.

Most methods that achieve this control are based on reverse engineering: one
imposes a given evolution for the system's probability density function and then
deduces the form of the driving protocol potential. This method is effective for
single-particle systems. Theoretically, one can replace the proposed probability
density function in the Fokker-Planck equation to obtain the form for
controlling the external potential. However, it is not clear how to extend this
method to many-particle systems. One can use the controlling potential deduced
for a single particle as a starting point for many interacting particles, but
this does not guarantee full control of all particles~\cite{Dago2020}. In this
work, we propose a method for numerically designing control protocols. We
perform stochastic numerical simulations of the system of interest and employ
machine learning techniques—in particular, automatic differentiation~\cite
{autodiff_review}—to evaluate how changes in control potential affect the
statistical state of the system and drive it to the desired final state. This
method has some similarities to the image generative diffusion
models~\cite{Sohl15,Ho2020} whose objective is to recreate a complex probability
distribution from a simple one (normal distribution), although the exact
implementation has some key differences. Similar techniques have been used to
design protocols that optimize the work done on the system~\cite{Engel2023}.
Here, however, we are interested in controlling the final state of the system. A similar goal in the realm of active matter has been explored in Ref.~\cite{Whitelam2024} using evolutionary algorithms, which offer a complementary approach to the present work.

The remainder of this article is organized as follows. In Sec.~\ref{sec:tech},
we explain in detail our numerical technique. Then, in
Sec.~\ref{sec:brownian-gaussian} we apply it to the control of a Brownian
particle under a variable harmonic potential. This serves as a test of our
method as its performance can be compared to analytical methods. In
Sec.~\ref{sec:brownian-nongauss}, we apply our method to achieve swift
state-to-state transformations of a Brownian particle when the control potential
is not harmonic, as an illustration of a situation where analytical methods are
more difficult to implement.

\section{Numerical design of control protocols}
\label{sec:tech}
To explain our method clearly, we consider a simple system consisting of a Brownian particle with position $X$ that experiences thermal fluctuations at temperature $T$ and is driven by a controlling force $F(X,\lambda(t))$, which depends on the position of the particle and a set of external, time-dependent parameters $\lambda(t)$. The particle is in a viscous fluid that exerts a drag force, characterized by a friction coefficient $\gamma$. We assume that the inertial timescale is much shorter than both the relaxation time and the duration of the controlling protocol; therefore, we operate in the overdamped regime. We describe the stochastic dynamics of the particle using a Langevin equation
\begin{equation}
  \frac{dX}{dt}=F(X,\lambda(t))/\gamma + \sqrt{2D}\xi(t),
\end{equation}
where $D=k_B T/\gamma$ is the diffusion coefficient, $k_B$ denotes the Boltzmann
constant, and we model $\xi(t)$ as a Gaussian white noise satisfying $\langle
\xi(t) \rangle=0$ and $\langle \xi(t)\xi(t') \rangle = \delta(t-t')$. In the
rest of this paper, we work with dimensionless quantities defined as follows. We
first identify a time scale $\tau$, which can represent the relaxation time of
the system (for a fixed $\lambda$), the duration of the controlling protocol, or
the total simulation time. Then, the \dr{time is expressed in dimensionless units as} $\hat{t}=t/\tau$, the position as $\hat{x}=x/\sqrt{D\tau}$, the gaussian noise
as $\hat{\xi}(\hat{t})=\sqrt{\tau} \xi(t)$, and the force as
$\hat{F}=F\sqrt{D\tau}/(k_B T)$. The forces derives from a potential energy:
$F=-\partial_x U(x,\lambda)$, and the \dr{potential energy is expressed in dimensionless units as}
$\hat{U}=U(x,\lambda)/(k_B T)$. With these definitions, the Langevin equation
reads
\begin{equation}
  \label{eq:langevin}
  \frac{d\hat{X}}{d\hat{t}}=\hat{F}(\hat{X},\hat{\lambda}(\hat{t})) + \sqrt{2}\hat{\xi}(\hat{t}).
\end{equation}
For simplicity, we drop the hat in what follows. An equivalent description of
the system is through the Fokker-Planck equation for the position probability
density function $P(x,t)$, which reads
\begin{equation}
  \label{eq:Fokker-Planck}
  \partial_t P(x,t) = -\partial_x \left[ F(x,\lambda(t)) P(x,t) \right] + \partial_x^2 P(x,t).
\end{equation}

We now state the problem we want to solve. Assume that the system starts at
$t=0$ in equilibrium with probability density
$P(x,0)=P^{\text{eq}}_i(x)=\frac{1}{Z_i}e^{-U(x,\lambda_i)}$ for an initial
control parameter $\lambda_i$. Here, $Z_i=\int e^{-U(x,\lambda_i)} dx$ is the
corresponding initial partition function. A control protocol $\lambda(t)$ of
duration $t_f$ is applied to the system and at $t=t_f$ we wish the system to be
in a new thermal equilibrium state corresponding to a final value $\lambda_f$ of
the controlling parameters:
$P^{\text{eq}}_{f}(x)=\frac{1}{Z_f}e^{-U(x,\lambda_f)}$ with $Z_f = \int
e^{-U(x,\lambda_f)} dx$. For an arbitrary control protocol $\lambda(t)$, even
when we set $\lambda(0^-)=\lambda_i$ and $\lambda(t_f^+)=\lambda_f$, the system
lags behind the instantaneous Boltzmann distribution $\propto
e^{-U(x,\lambda(t))}$. Consequently, at $t=t_f$, the system is not in thermal
equilibrium and requires additional time to relax to the new equilibrium state.
Our goal is to design the protocol $\lambda(t)$ so that $P(x,t_f)$ approximates
the desired equilibrium $P^{\text{eq}}_{f}(x)=\frac{1}{Z_f}e^{-U(x,\lambda_f)}$
as closely as possible. We base our method on numerical simulations of the
system and on systematic adjustments to the control protocol.

\subsection{Numerical training algorithm}

To achieve this, we begin with an arbitrary protocol $\lambda(t)$. For example,
we choose a linear interpolation between $\lambda_i$ and $\lambda_f$, possibly
adding some random values. We perform $N_s\sim 100\,000$ parallel simulations of
the Langevin equation, Eq.~\eqref{eq:langevin}, on a graphics processing unit
(GPU) using the initial protocol $\lambda(t)$. These simulations produce a large
ensemble of trajectories $\{x_{n}(t)\}_{n\in\{1,\ldots,N_s\}}$ from which we
compute the final probability density function $P(x,t_f)$. We then compare the
calculated $P(x,t_f)$ with the desired equilibrium distribution
$P^{\text{eq}}_{f}(x)$. Initially, these two distributions differ significantly.
We quantify the difference between $P(x,t_f)$ and $P_f^{\text{eq}}(x)$ using a
{\em loss} function, which we will define later on. Thus, we will need adjust
the protocol $\lambda(t)$ to minimize the loss function. However, it will be
very inefficient to blindly do arbitrary changes to $\lambda(t)$ and rerun the
simulations to find how the final distribution has changed. To do this
effectively, we will use techniques from machine learning, in particular
automatic differentiation~\cite{autodiff_review}, which allows us to know in
advance in which directions the changes on the protocol will make smaller the
loss function between the final distribution and the desired one. 

Although automatic differentiation is a well-known technique in the realm of
machine learning, we briefly review its key concepts to ensure completeness and
clarify the technical challenges we overcame. In a single pass of the system
simulation, the final probability density function $P(x,t_f)$ and the loss
function result from numerous elementary operations, such as sums and multiplications, performed throughout. Although we cannot analytically compute
$P(x,t_f)$, the loss function, or its gradient with respect to $\lambda$, we
know the partial derivatives of each elementary operation with respect to its
inputs. Throughout the simulation, these operations form a computational tree.
We store the partial derivatives of each elementary operation along with each
intermediate result. Then, the computational tree is traced backward and the
chain rule is applied to compute the gradient of the loss function with respect
to the control protocol $\lambda$. This technique is known as backpropagation in
automatic differentiation. 

\begin{figure}
  \centering
  \includegraphics[width=0.99\textwidth]
  {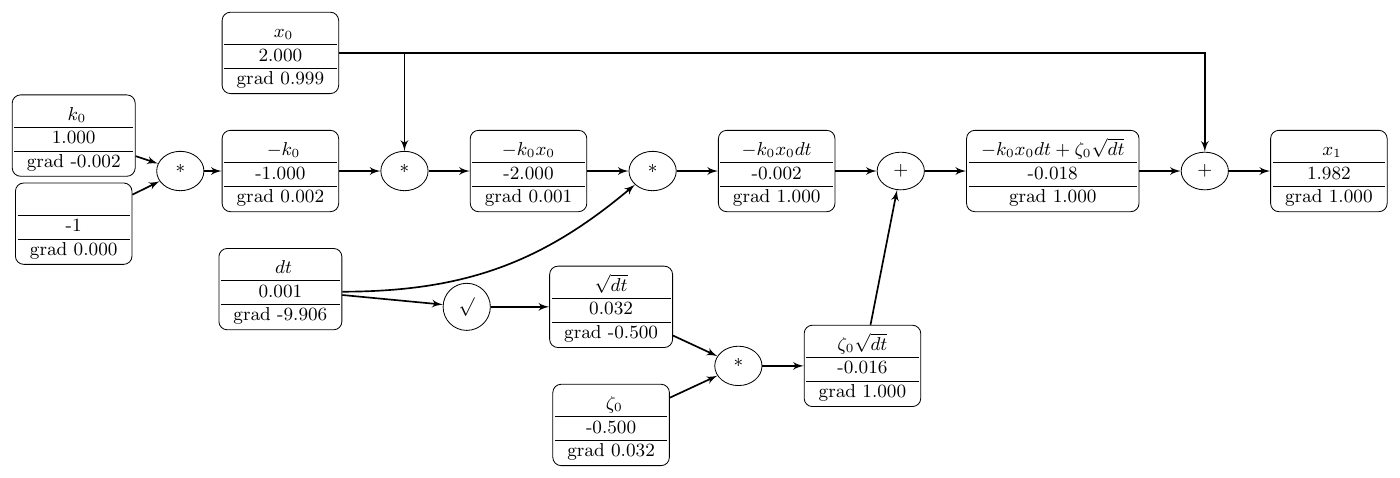} \caption{Computational graph for a single step of the Langevin
  equation, Eq.~\eqref{eq:euler-maruyama}, and its corresponding gradients. In
  this example, we used numerical values $x_0=2$, $k_0=1$, $dt=0.001$, and
  $\zeta_0=-0.5$ (a particular realization drawn from a normal distribution
  ${\cal N}(0,\sqrt{2})$). The graph is composed of boxes (rectangles) and
  operations (ellipses). Each box has three subdivisions indicating the
  variable, its value, and the gradient of $x_1$ with respect to the variable in
  the actual box, respectively.}
  \label{fig:step_graph_grad}
\end{figure}

For example, a single step of the resolution of the Langevin equation, using the
Euler-Maruyama method, in the Itô discretization, can be written as
\begin{equation}
  \label{eq:euler-maruyama}
  x_1= x_0 - k_0 x_0 dt + \zeta_0 \sqrt{dt}
\end{equation}
for a force $F=-k_0 x$ and where $\zeta_0\sim{\cal N}(0,\sqrt{2})$. The
particle moves from an initial position $x_0$ to a position $x_1$ in a time
$dt$. Figure~\ref{fig:step_graph_grad} shows the computational tree for this
operation. From the initial set of parameters ($x_0$, $k_0$, etc.), the values
are calculated forward, and the gradients are calculated backward. 
Suppose that the stiffness $k_0$ is one of the adjustable
parameters. In Fig.~\ref{fig:step_graph_grad}, we choose $k_0=1$, $x_0=2$ and
$dt=0.001$ as inputs. Multiplications, sums and square root operations are
performed as shown in the computational graph to obtain the value of the output
$x_1=1.982$. To compute the gradient of the output of
Eq.~(\ref{eq:euler-maruyama}), $x_1$, with respect to $k_0$, $\frac{\partial
x_1}{\partial k_0}$, we analyze how changes in $k_0$ propagate through the
computational tree in the backward direction. We start in the last box of the
computational tree in Fig.~\ref{fig:step_graph_grad} where the value of $x_1$ is
reported and the gradient $\partial x_1/\partial x_1=1$ is stored. The previous
operation is a sum. The partial derivative of the output of the sum with respect
to each of its inputs is $1$, so the gradient of $x_1$ with respect to
$-k_0x_0dt+\zeta_0\sqrt{dt}$ is $1$ as shown in the preceding box. 

Continuing the backward pass, we encounter another sum, again with a gradient of
$1$. We arrive at a multiplication between $dt$ and $-k_0x_0$. The partial
derivative of the result of this multiplication with respect to $-k_0 x_0$ is
simply $dt$, the other factor in the multiplication. In this example, we choose
$dt=0.001$, therefore the gradient of $x_1$ with respect to $-k_0x_0$ is now
$0.001$ as shown in the box labeled $-k_0 x_0$. Continuing backwards, we reach a
multiplication between $-k_0$ by $x_0$. The partial derivative $-k_0 x_0$ with
respect to $-k_0$ is $x_0=2$. Applying the chain rule, we multiply this partial
derivative, $2$, with the accumulated gradient, $0.001$, to obtain the gradient of
$x_1$ with respect to $-k_0$: $2\times0.001=0.002$. The last operation is a
multiplication of $-1$ by $k_0$, which finally gives the desired
gradient $\partial x_1/\partial k_0 = -0.002$. The backpropagation process
efficiently tracks these contributions by multiplying local derivatives along
the computational path, demonstrating how automatic differentiation captures the
sensitivity of $x_1$ to small changes in $k_0$. This method is significantly
more efficient than naive finite differences, which require recomputing $x_1$ for
each perturbation of $k_0$. 

In this previous example, $k_0$ appears only in one branch of the computational tree
which simplifies the analysis of the backward propagation of the gradient. When
a parameter appears on more branches, one needs to add each gradient
contribution when they merge. For example, in the $x_0$ box, we obtain $\partial
x_1/\partial x_0=0.999$ by adding the gradient contributions from two branches:
$1$ and $-0.001$.

For a review of automatic differentiation, see Ref.~\cite{autodiff_review}; an
excellent tutorial appears in \cite{micrograd}. In practice, we did not
implement automatic differentiation ourselves; instead, we used an existing
machine learning library optimized for parallel GPU computing. We used
\texttt{pytorch} \cite{pytorch}, however our algorithm could also be easily ported to
other frameworks such as \texttt{tensorflow} \cite{tensorflow} or
\texttt{jax} \cite{jax}.

After making the necessary changes in the control protocol $\lambda(t)$ in the
direction provided by the gradients (gradient descent method), the algorithm
starts over to perform a new set of simulations, evaluates again the loss
function, and adjusts the protocol until the loss function converges to the
smallest possible value. Each pass of simulations, computing the gradients and
adjusting the protocol is called an epoch following the machine learning
terminology.

This algorithm is very similar to the
one on which neural network training is based; therefore, the use of a
dedicated library such as \texttt{pytorch} is very convenient. In a sense, it is a
particular form of a neural network composed of $N_T=t_f/dt$ layers, each of
which performs a single step in the simulation:
\begin{equation}
  x(t+dt) = x(t) + F(x(t),\lambda(t)) dt + \sqrt{2} dW(t)
  \label{eq:langevin-discrete}
\end{equation}
with $W(t)$ a Wiener process. The network parameters are encoded in
$\lambda(t)$. This neural network takes as input the initial samples
$\{x_n(0)\}_{n\in\{1,\ldots,N_s\}}$ sampled from a Boltzmann distribution
$e^{-U(x,\lambda_i)}/Z_i$. In parallel, each sample goes through the layers
performing the calculations given by Eq.~\eqref{eq:langevin-discrete} to obtain
after $N_T$ layers the final samples $\{x_n(t_f)\}_{n\in\{1,\ldots,N_s\}}$. Note
that, at this stage, each sample is independent of the rest. Then a final layer
takes all the samples and combines them to compute the loss function $L$ (see
Sec.~\ref{sec:loss} for details). The gradients $\partial_{\lambda(t)} L$ of
the loss function with respect to the parameters $\lambda(t)$ are computed using
backpropagation, and the parameters are updated accordingly:
\begin{equation}
  \lambda(t) \leftarrow \lambda(t) - \eta \partial_{\lambda(t)} L,
  \label{eq:gradient-descent-general}
\end{equation}
where $\eta$ is the learning rate. The precise meaning of
$\partial_{\lambda(t)}L$ and Eq.~\eqref{eq:gradient-descent-general} depends on
the way the protocol is parametrized, see Sec.~\ref{sec:parametrization} for
details. This process is repeated until the loss function converges to a
minimum.

\subsection{Choice of the loss function}
\label{sec:loss}

Let us now discuss in more detail some important technical aspects of the
algorithm, starting with the choice of the loss function. A first natural choice
would be to use the Kullback-Leibler divergence 
\begin{equation}
  D(P(x,t_f)||P_f^{\text{eq}})= \int P(x,t_f) 
  \ln \left[\frac{P(x,t_f)}{P_f^{\text{eq}}(x)}
  \right]
  \,dx.
\end{equation}
This will require computing $P(x,t_f)$ from the set of samples $\{x(t_f)\}$
obtained from the simulation at the final time $t_f$. Numerically, this could be
done by building a histogram. Unfortunately, creating histograms requires selecting bin sizes, a step that is not differentiable and lacks support for automatic differentiation. For example, in \texttt{pytorch}, the \texttt{torch.histogram}
function does not have a backward method implemented to compute the gradients.
Therefore, a different choice of loss function is needed.

For the case where the force exerted on the particle derives from a harmonic
potential (Sec.~\ref{sec:brownian-gaussian}), one can show that if the initial
probability distribution is Gaussian, it will remain Gaussian at all times. In
this case a simple loss function can be chosen by comparing the average value of
the position 
\begin{equation}
  \label{eq:loss-mean}
  L_{\text{mean}} = \left(\left\langle x(t_f) \right\rangle - \left\langle x \right\rangle^{\text{eq}}_f\right)^2,
\end{equation}
and/or its variance
\begin{equation}
  \label{eq:loss-variance}
  L_{\text{var}} = \left(\left\langle \Delta x(t_f)^2 \right\rangle - \left\langle \Delta x^2 \right\rangle^{\text{eq}}_f\right)^2.
\end{equation}
Here, the average 
\begin{align}
  \left\langle x(t_f) \right\rangle &=
  \int x P(x,t_f) dx \\
  &\simeq
  \frac{1}{N_s} \sum_{n=1}^{N_s} x_n(t_f) 
\end{align}
is computed from the set of samples of the simulation, and the equilibrium one 
\begin{equation}
  \left\langle x \right\rangle^{\text{eq}}_f = \int x P^{\text{eq}}_f(x) dx
\end{equation}
is known directly from the analytic expression of $P^{\text{eq}}_f(x)$. For the
variance, we have defined $\Delta x(t_f)=x(t_f)-\left\langle x(t_f)
\right\rangle$ and $\Delta x=x-\left\langle x \right\rangle^{\text{eq}}_f$.
These loss functions defined in Eqs.~\eqref{eq:loss-mean} and
\eqref{eq:loss-variance} are differentiable and can be used to adjust the
control protocol.

However, beyond the harmonic case, it is not enough
to verify that the final average and variance converge to the expected values. 
One could compare higher order cumulants, or better yet, compare the
characteristic functions of the final distribution 
\begin{equation}
  \tilde{P}(k,t_f) = \left\langle e^{ikx(t_f)} \right\rangle
\end{equation}
with the expected one
\begin{equation}
  \tilde{P}^{\text{eq}}_{f}(k) = \int P^{\text{eq}}_{f}(x) \, e^{ikx} dx.
\end{equation}
Contrary to the non-differentiable computation of $P(x,t_f)$ by building a histogram, the
computation of the characteristic function is a differentiable operation from
the samples $\{x_{n}(t_f)\}_{n\in\{1,\ldots,N_s\}}$ as it involves computing the
exponential function and summing over
\begin{equation}
  \tilde{P}(k,t_f) \simeq \frac{1}{N_s} \sum_{n=1}^{N_s} e^{ikx_n(t_f)}.
\end{equation}
For this more general case, the loss function is 
\begin{equation}
  \label{eq:loss-char}
  L_{\text{char}} = \int \left| \tilde{P}(k,t_f) - \tilde{P}^{\text{eq}}_{f}(k) \right|^2 dk.
\end{equation}
Numerically, this is approximated by choosing an appropriate range of values for
$k$ and replacing the integral by a sum over those values of $k$. 

\subsection{Parametrization of the protocol}
\label{sec:parametrization}

The control protocol $\lambda(t)$ can be parametrized in many ways. For example,
$\lambda(t)$ can be a piecewise linear function, a polynomial, a rational
function, etc. The choice of the parametrization will depend on the specific
problem at hand. For example, if the control protocol is expected to be smooth,
a polynomial or a rational function parametrization could be a good choice. In
our examples, we explore a few possibilities: a piecewise linear
parameterization, a polynomial parametrization, and a neural network to
approximate $\lambda(t)$. 

The piecewise linear parametrization is the most intuitive one. The duration of the protocol
$[0,t_f]$ is divided into $N_p$ time intervals of equal duration:
$0=t_0<t_1<\ldots<t_{N_p}=t_f$, with $t_j=j\Delta t$ where $\Delta t=t_f/N_p$.
The value of $\lambda(t_j)=\lambda_j$ at the beginning of each interval is a
parameter of the protocol. Between times $t_j$ and $t_{j+1}$, $\lambda(t)$ is a
linear interpolation between $\lambda_j$ and $\lambda_{j+1}$. The total number
of parameters is $N_p$. The protocol is then defined by the set of parameters
$\{\lambda_1,\ldots,\lambda_{N_p}\}$ that are the variables to be adjusted to
minimize the loss function. After each simulation, the gradients of the
loss function $L$ with respect to the parameters $\lambda_j$ are calculated and
the new values of the parameters are updated accordingly to $\lambda_j
\leftarrow \lambda_j - \eta \partial_{\lambda_j} L$ where $\eta$ is the learning
rate.

In a polynomial parametrization, $\lambda(t)$ is a polynomial function of time:
\begin{equation}\label{eq:polynomial-parametrization}
  \lambda(t) = \sum_{n=0}^{P} a_n t^n.
\end{equation}
The coefficients $a_n$ are the variables that the algorithm will optimize. After
each simulation, these are updated using gradient descent:
$a_n\leftarrow a_n - \eta \partial_{a_n} L$.

Finally, based on the universal approximation theorem~\cite{Hornik89}, we also
explored a neural network parametrization of $\lambda(t)$. 
We selected a feedforward neural network architecture, which can be implemented
using \texttt{PyTorch}. The input layer consists of a single neuron representing
time, $t$. This neuron processes $t$ through an activation function and passes
the result to the next layer, which contains $n$ neurons. Each neuron in this
layer applies an activation function to the input it receives and forwards the
output to subsequent layers. This process continues until the final layer whose
output is of dimension one determining the output of the model $\lambda(t)$. An
example of this neural network parametrization of the protocol is shown in
subsection~\ref{subsubsec:work-neuralnet}.

\subsection{Relation with previous works}

In recent years, there has been a growing interest in applying machine learning techniques in statistical physics, from the control and design of self-assembling materials~\cite{whitelamLearningGrowControl2020,goodrichDesigningSelfassemblingKinetics2021}, train feedback protocols (Maxwell demons) to extract work and absorb entropy~\cite{Whitelam2023}, do memory erasure~\cite{whitelamHowTrainYour2023,barrosLearningEfficientErasure2025}, and to optimize the control of non-equilibrium systems~\cite{Engel2023,Loos2024, Whitelam2024}. This interplay has been fruitful both ways, as concepts from statistical physics have open the way for new techniques in machine learning, for example in diffusion generative models~\cite{Sohl15, Ho2020,biroliGenerativeDiffusionVery2023,biroliDynamicalRegimesDiffusion2024a} or training neural networks using Monte Carlo methods~\cite{Whitelam_2022}.

Our method is inspired by previous work~\cite{Engel2023} on optimal control
of non-equilibrium systems. In that work, stochastic simulations of a system of
interest are performed in a differentiable way to compute the gradients of a
loss function with respect to the control protocol. The loss function used in
Ref.~\cite{Engel2023} was the entropy production or the work done on the system.
Our goal here is different; the loss function we use is to minimize
the difference between the final probability density function of the system and
the desired equilibrium distribution. Our algorithm implementation is
independent from~\cite{Engel2023} using \texttt{pytorch} instead of
\texttt{jax}.

The optimal control of stochastic systems and driving them from an initial state to a target final state can also be solved by deterministic optimal transport techniques as shown in Ref.~\cite{Optimal_Protocols_Optimal_Transport}, where the stochastic thermodynamics optimization of heat or work  problem is mapped into a Burgers equation. Then standard optimal transport techniques~\cite{villani2003topicsTransportation} can be used to find optimal control protocols~\cite{Baldovin2025,klingerUniversalEnergyspeedaccuracyTradeoffs2025,muratore-ginanneschiHowNanomechanicalSystems2014}. Our present work is an alternative to those methods.

In the realm of machine learning, there has been recent interest in neural
ordinary differential equations~\cite{chenNeuralOrdinaryDifferential2019a},
where the dynamics $\{x(t)\}$ of a system is given by an ordinary differential
equation 
\begin{equation}
  \frac{dx}{dt} = f(x(t),t,\lambda),
\end{equation}
and $f$ is described by a neural network, with parameters $\lambda$ that are to
be optimized such that the system is in a desired final state $x(t_f)$. In
Ref.~\cite{chenNeuralOrdinaryDifferential2019a}, a method was devised to compute the
gradients of the loss function without backpropagating through all the
intermediate $N_T$ steps. The method is known as the adjoint sensitivity
method~\cite{Pontryagin62} and is quite memory efficient of $\mathcal{O}(1)$
instead of $\mathcal{O}(N_T)$ for the backpropagation method. Unfortunately, it is
not clear how to apply this method to stochastic differential equations such as
the one we are solving. However, a promising proposal to extend the adjoint
sensitivity method to deal with gradient computations for stochastic
differential equations in a memory efficient manner of order $\mathcal{O}(1)$
was proposed in Ref.~\cite{liScalableGradientsStochastic2020}, at the expense of
a longer time computation $\mathcal{O}(N_T\log N_T)$. However, we did not need
to implement this method, as the backpropagation method through all intermediate computations was enough for our purposes. Since our objective is to find a control protocol to achieve fast thermalization in a short time
duration, the number of steps $N_T$ is not very large, and the memory bottleneck
was not reached in our simulations.

\section{Brownian particle under harmonic potentials}
\label{sec:brownian-gaussian}

In this section, we examine an overdamped Brownian particle subject to a
harmonic potential, the general expression for which is
\begin{equation}\label{eq: general harmonic potential}
  U(x,\lambda(t)) = \frac{1}{2} k(t)(x-c(t))^2.
\end{equation}
The external control parameter $\lambda(t)$ can represent here either the potential
center $c(t)$ or the stiffness $k(t)$, or both. The core concept of optimal
control involves determining a form of $\lambda$ that allows the system to
optimize or reduce a specific desired property. In the following, the focus is
on control protocols that speed up the equilibration time or optimize the work
performed on the particle, or aim at both goals simultaneously.

\subsection{Changing the stiffness of the harmonic potential}

First, we analyze a system with a fixed center, $c(t) = 0$, and a time-dependent
stiffness, $k(t)$, subject to the boundary conditions $k(0^{-}) =
k_i$ and $k(t_f^{+}) = k_f$. We express the system in dimensionless units based on
the natural relaxation time $\tau = \gamma / k_f$, which depends on the final
stiffness $k_f$, as outlined in Sec.~\ref{sec:tech}. This is the relaxation time
if an abrupt quench, where the potential transforms instantaneously from the
initial to the final potential is performed. We seek a control protocol that
satisfies the boundary conditions while ensuring that $t_f < \tau$.

Researchers have proposed several analytical protocols to accelerate thermal
equilibration in this
system~\cite{trizac1,guery-odelinDrivingRapidlyRemaining2023,RengifoTellez24}.
These protocols provide a theoretical framework for optimizing control
strategies that reduce relaxation times beyond what passive equilibration
allows. Thus, replicating these well-known results provides a valuable starting
point to validate our approach before applying it to more complex problems.

\subsubsection{One parameter protocol: the two-step protocol}

The simplest and most intuitive protocol is likely the two-step
protocol (TSP) \cite{RengifoTellez24}, where the stiffness changes abruptly from $k_i$
to some value $k_m$ at $t=0$ and then from $k_m$ to $k_f$ at $t=t_f$. The value
of $k_m$ is increased sufficiently to ensure that the particle relaxes faster to
the new equilibrium state at $t_f$. 

Using the Fokker-Planck equation~\eqref{eq:Fokker-Planck} with the potential
Eq.~\eqref{eq: general harmonic potential}, one can obtain the evolution
equation for all cumulants $\chi_m$ of the probability density function of the
position. From there, one can show that if the system starts with an initial
probability density function that is Gaussian and the driving protocol is
harmonic, all cumulants except the second one (the variance) will be null at all
times. This means that the probability density function will remain Gaussian at
all times, with a time-dependent variance $\chi_2(t)$ satisfying  
\begin{equation}
  \label{eq:chi2}
  \frac{d\chi_2(t)}{dt} + 2k(t) \chi_2(t) = 2.
\end{equation} 
To achieve thermalization, the variance at the final time $t_f$ must be
$\chi_2(t_f)=1/k_f$ so that $P(x,t_f)=P^{\text{eq}}_f(x)=e^{-k_f
x^2/2}/\sqrt{2\pi/k_f}$. Since the stiffness is constant in the time interval
$(0,t_f)$, $x(t)$ is an Ornstein-Uhlenbeck process whose probability density
function $P(x,t)$ is a Gaussian with zero mean and variance
\begin{equation}
  \chi_2(t)=\frac{1}{k_m}+\left(\frac{1}{k_i}-\frac{1}{k_m}\right)e^{-2k_m t}.
\end{equation}
Using $\chi_2(t_f)=1/k_f$, we obtain the following transcendental equation for
$k_m$ (for details see Ref.~\cite{RengifoTellez24})
\begin{equation}
  \label{eq:TSP-k_m}
  \frac{1}{k_f}=\frac{1}{k_m}+\left( 
    \frac{1}{k_i}-\frac{1}{k_m}
  \right)e^{-2k_m t_f}.
\end{equation} 
This analytic solution of the problem can be used as a test of our numerical
training algorithm. We use a parametrization of $\kappa(t)$ with a constant
value $k_m$ between $(0,t_f)$ and discontinuous at $t=0$ and $t=t_f$. This
protocol has only one adjustable parameter which is $k_m$. The loss function
can be either the variance loss function, Eq.~(\ref{eq:loss-variance}), or the
characteristic function loss function, Eq.~(\ref{eq:loss-char}). 

\begin{figure}[ht]
  \centering


\includegraphics[width=1.0\textwidth]{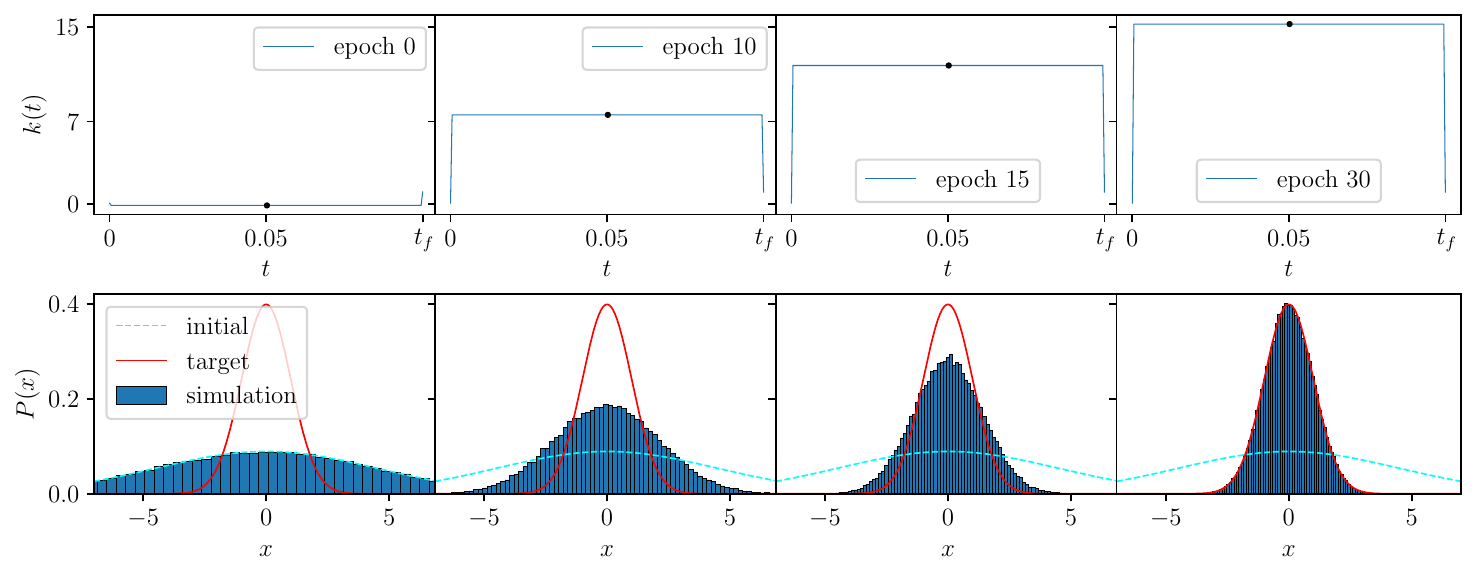}
  \caption{ Training progress over 30 epochs for the two-step protocol applied to the stiffness parameter, utilizing the loss function defined in Eq.~\eqref{eq:loss-char} and a learning rate of $\eta=100$. The initial (dashed) and target (solid) probability distributions are displayed, alongside a histogram generated from the model trained up to the specified epoch. The system parameters are set to $k_i=0.05$, $k_f=1.0$ and $t_f=0.1$.
  \label{fig:TSP}}

\end{figure}

For numerical calculations, we used the values $k_i=0.05$ and $k_f=1.0$, for
which the relaxation time $\tau=1/k_f=1.0$. We aim to
achieve equilibration at time $t_f=0.1$, an order of magnitude faster than the
natural relaxation. We trained this one-parameter model during 30 epochs with a
learning rate of $\eta=100.0$. This large value (compared to the traditional
values used in machine learning of order $10^{-2}$) is understood because there
is only one parameter to adjust. For training, we used the characteristic
function loss function in Eq.~\eqref{eq:loss-char}. In Fig.~\ref{fig:TSP}, we
show the training results for the stiffness as a function of time after some epochs.
This plot also includes the probability density function (PDF) of the position,
at the initial time, the target PDF, and the position histogram after simulating
the system with the protocol given by the respective epoch. During the first 10
epochs, the system lags behind the final equilibrium distribution at $t_f$. But
after epoch 10, the adjustment made to $k_m$ drives the system faster and
closer to the expected final equilibrium distribution. By epoch 30, the final
probability density is indistinguishable from the expected one, where the loss
function was of order $10^{-5}$. The value obtained from $k_m$ with this
training was $k_m=15.27$, while the theoretical value from
Eq.~\eqref{eq:TSP-k_m} is $k_m=15.30$. This corresponds to a precision of
$0.20\%$ relative to the expected value.

\subsubsection{Linear interpolation protocols}

\begin{figure}
  \includegraphics[width=0.5\textwidth]{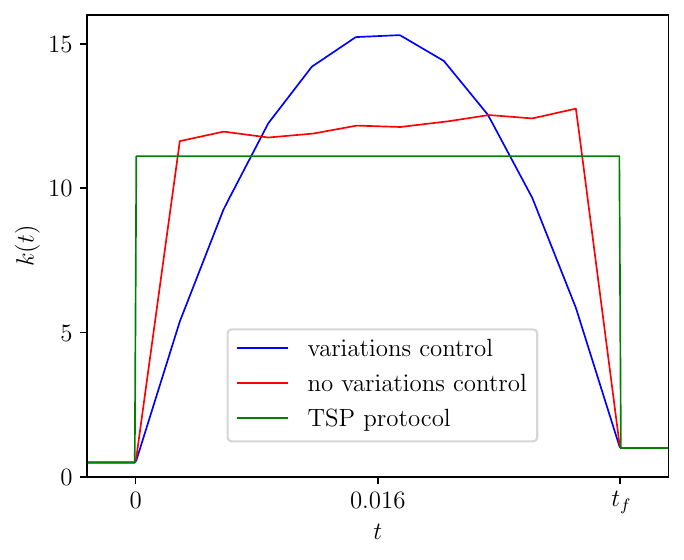}
  \caption{Stiffness as a function of time designed to accelerate equilibration. The red curve represents a 10-parameter protocol trained using the characteristic function loss, Eq.~\eqref{eq:loss-char}. The blue curve also corresponds to a 10-parameter protocol but includes an additional penalty term in the loss function to suppress large variations in stiffness, as given in Eq.~\eqref{eq:loss_FT_and_work}. For reference, the green curve depicts the corresponding two-step protocol.  The initial and final stiffness values are set to $k_i = 0.5$ and $k_f = 1$, respectively, with a protocol duration of $t_f = 1/30$.\label{fig:kappa_comparison}}
\end{figure}

We now consider a more general case where the protocol $k(t)$ is not constant in the interval $(0,t_f)$ and is continuous at the boundary. In this case, we can parameterize $k$ using piecewise linear parameterization with $N_p=10$ parameters. It is important to recall that the time interval $[0,t_f]$ is divided into $N_p$ subintervals and the adjustable parameters are the values of $k(t_j)$ at the beginning of each subinterval as
explained in Sect.~\ref{sec:parametrization}. For the training process, we took the initial and final stiffness to be $k_i=0.5$
and $k_f=1$, respectively. These values lead to a relaxation time $\tau=1$ and we set the duration of the protocol to be $t_f=1/30$. Figure~\ref{fig:kappa_comparison} shows results for two trained protocols and the TSP for reference. The protocol represented by the red curve was obtained through optimization with the characteristic function loss Eq.~\eqref{eq:loss-char}, during 60 epochs. Notice that this protocol has a very fast variation at the initial $t_i=0$ and final times
$t_f$, and does not change much during the rest of the time interval. It
quite resembles the corresponding two-step protocol shown in green. There are an infinite number of solutions to the thermalization
problem; nevertheless, this shows that in the training algorithm the two-step
protocol plays an important role as it exhibits a strong and stable attraction
for the algorithm. 

In a practical implementation of the protocol, it might not
be desirable to have very large variations in the stiffness in a short time or
discontinuities in it. To achieve a smoother protocol, we controlled the
variations of $k(t)$ between successive time intervals by adding to the
loss function a term penalizing strong variations:
\begin{equation}\label{eq:loss_k_varitaions}
  L_{\text{grad$k$}} = \sum_{j=0}^{N_p-1} 
  \left(k(t_{j+1})-k(t_j) \right)^2.
\end{equation}
Then, as a loss function, we can adopt the combination $L=(1-\alpha)L_{\text{char}}+\alpha L_{\text{grad$k$}}$, where
$\alpha$ is a mixture parameter which we initially fixed at $\alpha=10^{-4}$ for 60 epochs,
then turned it off at $\alpha=0$ for another 60 epochs of training. This procedure
results in much smoother stiffness as shown in blue in
Fig.~\ref{fig:kappa_comparison}. Introducing restrictions or desirable properties into the loss function is a crucial tool, as it enables the customization of control protocols to exhibit convenient and tailored properties.

It can be observed that all three protocols in Fig.~\ref{fig:kappa_comparison}
require a large increase in the mean value of $k(t)$ to shorten the
effective relaxation time of the protocol. This is a common feature of all
protocols that has been analyzed in Ref.~\cite{RengifoTellez24}. The average
value of $k(t)$ during the duration of the protocol is of the order
$\frac{1}{2t_f}\ln\frac{k_f}{k_i}$ for $t_f\ll 1$.

\subsubsection{Polynomial parametrization}

The piecewise linear parametrization model might initially seem restrictive due to its reliance on linear interpolation. However, during the optimization of the loss function, non-linearities are inherently accounted for. In addition, increasing the number of parameters can improve the model's ability to reproduce more complex protocols. Alternatively, models can be defined to incorporate non-linearities from the outset. For instance, in our system, the stiffness can be parameterized using a polynomial, with its coefficients serving as trainable parameters.

\begin{figure}[h]
  \centering
  \includegraphics[width=0.5\textwidth]{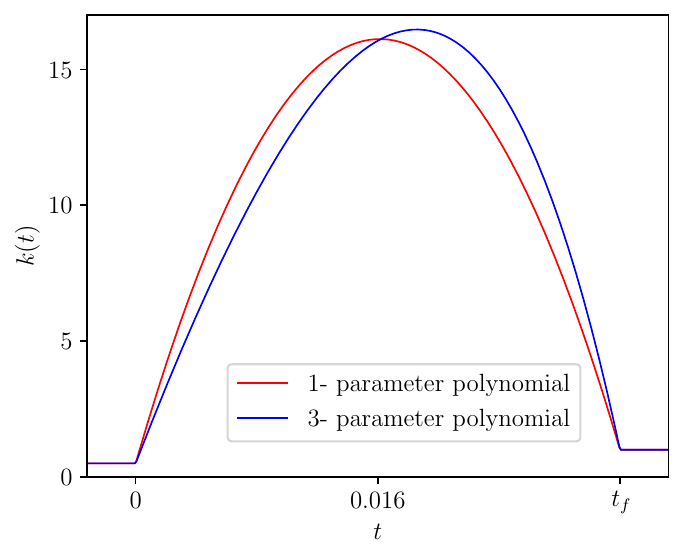} 
  \caption{Results for stiffness $k(t)$ after training two models using the polynomial parameterization defined in Eq.~\eqref{eq:kappa_polynomial}. The red curve represents the model trained with a single parameter, while the blue curve corresponds to the model trained with three parameters. The initial and final stiffness values are $k_i = 0.5$ and $k_f = 1$, respectively, with a protocol duration of $t_f = 1/30$.}
  \label{fig:kappa_polynomial}
\end{figure}

Consider a polynomial parametrization for the stiffness given by 
\begin{equation}
  \label{eq:kappa_polynomial}
  k(t) = k_i + s (k_f - k_i) + s (1-s) p(s)
\end{equation}
where $s = t/t_f$, and $p(s)$ is a polynomial
\begin{equation}
  p(s) = \sum_{n=0}^{N_p-1} a_n s^n.
\end{equation}
The coefficients $a_n$ are the parameters that the algorithm will optimize. The
form proposed for the stiffness, Eq.~\eqref{eq:kappa_polynomial}, ensures
continuity at $t=0$ and $t=t_f$. Figure~\ref{fig:kappa_polynomial} shows two
trained protocols, one with a single coefficient (1-parameter, $p$ has degree 0)
and the other with three coefficients (3-parameters, $p$ has degree 2). The
initial and final stiffnesses are $k_i=0.5$ and $k_f=1$, and the duration of the
protocol is $t_f=1/30$. Both protocols were trained for 60 epochs and the loss
function used was the characteristic loss function Eq.~\eqref{eq:loss-char},
which at the end of training was of the order $10^{-5}$. For the one-parameter
protocol, the parameter value was $a_0=61.69$, and for the three-parameter
protocol: $a_0=45.84$, $a_1 =23.51$, and $a_2=14.54$. 

Although the number of parameters $N_p$ can be chosen arbitrarily, for this
particular problem, with a harmonic potential, one can restrict the search of a
solution to a one-dimensional space. This is so because the probability density function
remain gaussian at all times, with a time-dependent variance $\chi_2(t)$ that
satisfies Eq.~(\ref{eq:chi2}). To achieve a fast thermalization, the variance at
the final time $t_f$ needs to be $1/k_f$. Therefore, one can parametrize
$\kappa(t)$ with only one parameter that get fixed by the condition
$\chi_2(t_f)=1/k_f$. However, for more complex potentials, such as the quartic potential discussed in Sec.~\ref{sec:brownian-nongauss}, the search space must be expanded, necessitating an increase in the number of parameters. This is because additional cumulants, $\chi_m$, become nonzero, and their evolution is governed by a hierarchy of coupled differential equations that must be solved simultaneously.

\subsection{Fast thermal protocol that optimize work}

\subsubsection{Optimizing work alone}
\label{subsubsec:work-neuralnet}

\begin{figure}[h]
  \centering 
  \includegraphics[width=0.5\textwidth]{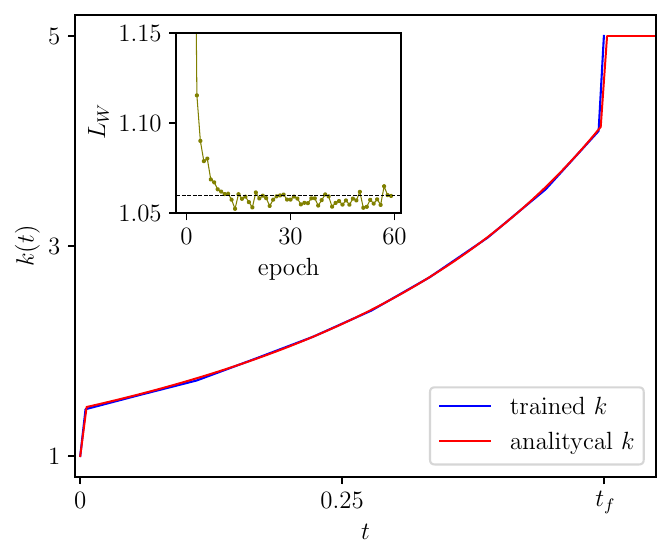}
  \caption{Stiffness as a function of time for the optimal work protocol (red curve), given by Eq.~\eqref{eq:optimal_protocol_work_stiffness}, and the trained protocol (blue curve) consisting of a piecewise linear interpolation with 10 learnable parameters. The initial and final stiffness values were $k_i = 1.0$ and $k_f = 5.0$, respectively, with a final time of $t_f = 1.0$ and a fixed center at $c = 0.0$. The inset depicts the evolution of the loss function during training, with the horizontal line representing the optimal work value, $\expval{W}_{opt} = 1.0565$.\label{fig:stiffness_protocol_opt_work}}
\end{figure}

In the previous examples, we focused on demonstrating how to train an algorithm capable of achieving fast thermalization. Now, we aim to show that the same algorithm also performs exceptionally well in identifying protocols that optimize the work done on the system. In the realm of stochastic thermodynamics \cite{peliti_book}, the work done on a system is a random variable described by
\begin{equation}
  W = \int_{0}^{t_f}  \dot{\lambda}(t) \frac{\partial U}{\partial\lambda}(x,\lambda(t))
  \,dt
  ,  
\end{equation}
where $\dot{\lambda}(t)$ represents the time derivative of the control parameter $\lambda(t)$.
The average work done on the system can be used as a loss function
\begin{equation}\label{eq:average_work}
 L_{W}= \expval{W} = \int_{0}^{t_f} dt \dot{\lambda}(t) \expval{\frac{\partial U}{\partial\lambda}(x,\lambda(t))}= \frac{1}{2}\int_{0}^{t_f} dt \dot{k}(t) \expval{x^2},
\end{equation}
where the average $\expval{\cdot}$ is taken over the initial distribution and noise realizations. The last equality is valid only for the harmonic oscillator with $\lambda(t)=k(t)$. 
The problem of determining the optimal work protocol was addressed in~\cite{Seifer-optimal}, where mathematical expressions were derived for the optimal stiffness and center protocols. For reference, we include their results here. The optimal stiffness protocol within the interval $0<t<t_f$ is given by:
\begin{equation}\label{eq:optimal_protocol_work_stiffness}
  k_{opt}(t) = \frac{k_i-A(1+At)}{(1+At)^2},
\end{equation}
and the optimal work is 
\begin{equation}\label{eq:optimal_work_stiffness}
    \expval{W}_{opt} =\frac{(At_f)^ 2}{k_i t_f}-\ln{(1+At_f)}+\frac{1}{2}\frac{k_f}{k_i}(1+At_f)^2 -\frac{1}{2},
\end{equation}
where the constant $A$ is  
\begin{equation}
  A = \frac{-1-k_f t_f + \sqrt{1+2k_i t_f + k_i k_f t^2_f}}{t_f(2+k_f t_f)}.
\end{equation}
To reproduce this result, a piecewise linear parametrization model with $N_p=10$ learnable parameters was trained over 60 epochs and a learning rate of $\eta=10.0$. For the numerical simulation, the parameters were $k_i=1.0$, $k_f=5.0$, and $t_f=1.0$. From equation~\eqref{eq:optimal_work_stiffness}, the expected work is \(\langle W \rangle_{opt} = 1.0565\). The value obtained after 60 epochs of training is \(\langle W \rangle = 1.0594\), corresponding to a relative error of $0.27\%$. This shows that the model accurately reproduces the optimal protocol for stiffness using average work as the loss function. The trained protocol and the analytical solution Eq.~\eqref{eq:optimal_protocol_work_stiffness} as well as the loss function are shown in Fig.~\ref{fig:stiffness_protocol_opt_work}.

\begin{figure}[h]
  \centering
    
    \includegraphics[width=0.8\textwidth]{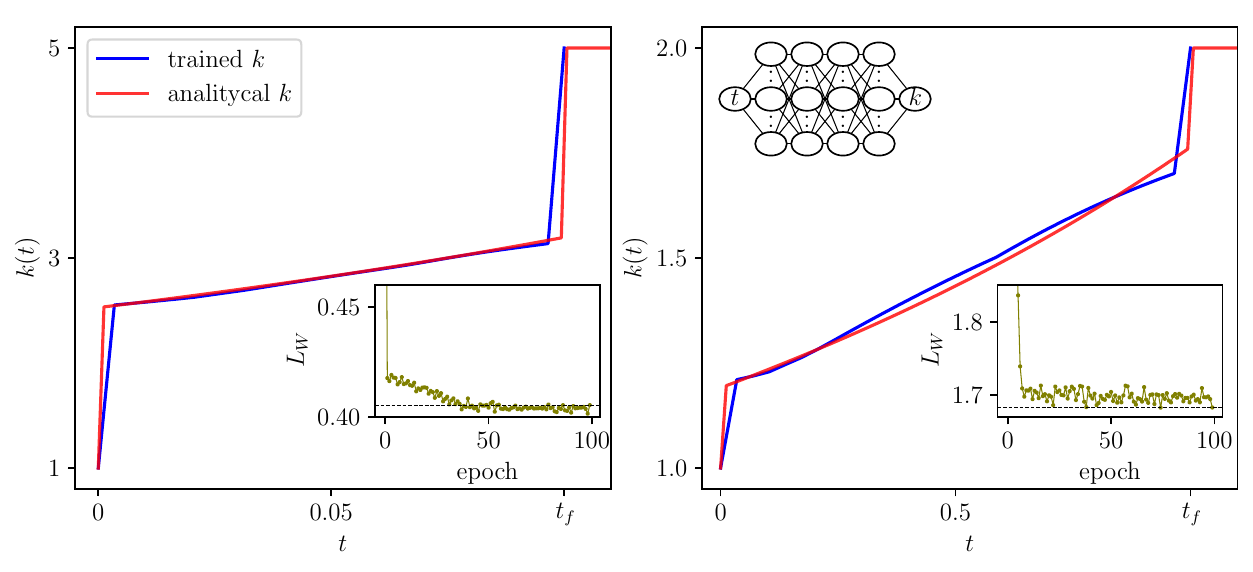}

  \caption{ Results after training a neural network to reproduce the stiffness that minimizes the work done on the system. The blue curve represents the trained model, while the red curve corresponds to the analytical solution given by Eq.~\eqref{eq:optimal_protocol_work_stiffness}. The insets display the evolution of the mean work (loss function) during training. The left panel corresponds to parameters \( k_i = 1.0 \), \( k_f = 5.0 \) and $t_f=0.1$ whereas the right panel corresponds to \( k_i = 1.0 \), \( k_f = 2.0 \), \( t_f = 1.0 \).\label{fig:NN_trained} }
  
\end{figure}

We can also work with parametrizations that involve neural networks. The protocol described in Eq.~\eqref{eq:optimal_protocol_work_stiffness} can be reproduced by a neural network whose architecture is as follows. The input $t$ is first normalized by dividing it by the time duration parameter $t_f$. The normalized input is then passed through the first fully connected layer, which maps it to a space of dimension $n=30$, followed by the application of the \texttt{ReLU} activation function. The output of this layer is fed into a second layer, where a \texttt{Tanh} activation is applied. A third layer follows, utilizing \texttt{Leaky ReLU} for handling small negative gradients. These deep layers have the same number of neurons $n=30$. The fourth and final layer produces a scalar output using a modified \texttt{Tanh} to constrain the result within the range of $k_i$ to $k_f$, see \footnote{The \texttt{Leaky ReLU} function is stepwise that returns $x$ if $x\geq 0$ and $-\epsilon x$ if $x<0$, where epsilon is a positive small value, by default in PyTorch, $\epsilon =10^{-2}$. If $\epsilon=0$, the function is called \texttt{ReLU}. The function \texttt{Tanh} is the usual hyperbolic tangent.}.
The neural network was trained for 100 epochs with a learning rate of $\eta=0.7$. The results are shown in Fig.~\ref{fig:NN_trained}. The left panel corresponds to the parameter set $k_i = 1.0$, $k_f = 5.0$ and $t_f=0.1$ from which the optimal work is \(\expval{ W}_{\text{opt}} = 1.6937 \) (Eq.~\eqref{eq:optimal_protocol_work_stiffness}). The final value of the loss function was \( L_{W} = 1.6945 \) representing a relative error of $0.04\%$. In the right panel of Fig.~\ref{fig:NN_trained}, the parameters were $k_i = 1.0$, \( k_f = 2.0 \), \( t_f = 1.0 \), with an optimal work of \(\expval{ W}_{\text{opt}} = 0.4029 \). The last value of the loss function was \( L_{W} = 0.4053 \) corresponding to a relative error of $0.60\%$. These results demonstrate that the network accurately reproduces the optimal stiffness protocol for both cases.


\subsubsection{Optimizing work and equilibration time}
Accelerating a system's equilibration through external control carries an energetic cost. Thus, a fundamental question is whether it is feasible to devise a control protocol that both accelerates equilibration and optimizes the work performed. This question holds significance not only from an experimental standpoint but also from a theoretical perspective. This issue has been solved in Ref.~\cite{Schmiedl_2008} where the optimal control protocol is the same Eq.~\eqref{eq:optimal_protocol_work_stiffness}; however, the constant $A$ now is given by
\begin{equation}\label{eq:second_A}
    A =\frac{1}{t_f} \left(\frac{k_i}{k_f}-1\right),
\end{equation} 
and the fast thermal (FT) optimal work is 
\begin{equation}\label{eq:work_FTW}
    \expval{W}_{opt} =\frac{1}{2}\ln(\frac{k_f}{k_i})+\frac{1}{k_i t_f}\left(\sqrt{\frac{k_i}{k_f}}-1\right)^2 .
\end{equation}
Observe that the quasi-static expression is reached if $t_f\rightarrow\infty$. See Ref.~\cite{Schmiedl_2008}.
The change in the constant $A$ is due to a modification of the boundary conditions that ensures that the final state is an equilibrium state, as required by rapid thermal equilibration. To replicate this protocol, we now require a loss function that encompasses the two conditions to be optimized. Consequently, we propose a loss function $L_{FTW}$ 
\begin{equation}\label{eq:loss_FT_and_work}
    L_{FTW} = (1-\alpha)L_{char} + \alpha L_{W}.
\end{equation}
This is a convex linear combination of $L_{char}$ and $L_{W}$ of Eqs.~\eqref{eq:loss-char} and \eqref{eq:average_work}. The blend parameter $\alpha$ works as a hyperparameter that is introduced by the user. The model used was $N_p=10$ piecewise linear parametrization. The training was more challenging than the previous ones because the minimization of $L_{FTW}$ represents a compromise between minimizing the work done on the system and reducing the distance to the target equilibrium state. Depending on the value of $\alpha$ one can give more importance to one of these objectives. The training was performed in batches of 30 epochs, with a variable learning rate of $\eta$ in the range $15000$ to $30000$ and a variable blending $\alpha$. We started with $\alpha=0$ to let the algorithm find a fast thermal equilibration protocol, regardless of the work done. From there, we evaluated the average work $L_W$ to be able to compare it with $L_{char}$ and choose an appropriate blend $\alpha$ so that both contributions to $L_{FTW}$ are weighted equally. This resulted in choosing $\alpha$ in the range from $10^{-5}$ to $10^{-4}$ for batches of 30 epochs and adjusting accordingly after each batch.

\begin{figure}
    \centering
    \includegraphics[width=0.5\linewidth]{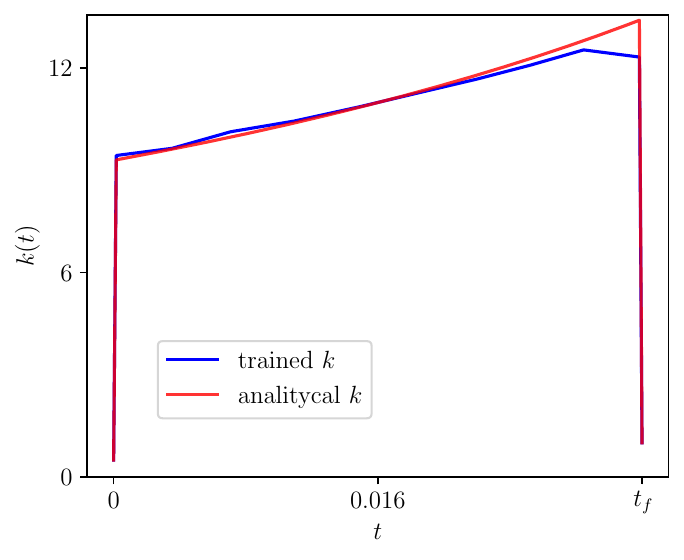}
    \caption{Stiffness $k$ as a function of time designed to accelerate
    equilibration and minimize work done on the system. The red curve represents
    the optimal protocol Eq.~\eqref{eq:optimal_protocol_work_stiffness} with $A$
    given by Eq.~\eqref{eq:second_A}, see Ref.~\cite{Schmiedl_2008}. Meanwhile,
    the blue curve illustrates the trained protocol. The parameters were $k_i =
    0.5$, $k_f = 1.0$, and $t_f = 1/30$.
    \label{fig:stiffness_protocol_opt_work_and_fastthermal}}
\end{figure}

Finally, we obtained an acceptable optimal protocol, as depicted in
Fig.~\ref{fig:stiffness_protocol_opt_work_and_fastthermal}. The theoretical
value provided by Eq.~\eqref{eq:work_FTW} is $\expval{W}_{opt}=5.4937$, whereas
the simulation yields a value of $L_W=\expval{W}=5.5030$. The relative error is
$0.17\%$, and the final loss function $L_{char}$ value was noted of
order $10^{-6}$, indicating that the equilibrium state has been reached. Note
that the trained protocol $k(t)$ in
Fig.~\ref{fig:stiffness_protocol_opt_work_and_fastthermal} is smaller from the
theoretical one in the last time steps. This shows the compromise mentioned
earlier between optimizing the work (smaller values of $k(t)$) and reaching the
equilibrium step at $t_f$ which requires larger values of $k(t)$.

\subsection{The center of the harmonic potential as a control parameter}

In the previous sections, we have focused on the case where the stiffness varies over time. However, the same protocol parametrization techniques apply when the stiffness is constant at $k = 1$, while the center position varies with time. To demonstrate that our machine learning approach is adaptable in this context as well, this section explores rapid thermal equilibration and work optimization by treating the center as a control parameter and employing a piecewise-linear parametrization.

The harmonic force in this case is
\begin{equation}\label{eq:force center}
F(x, \lambda(t)) = -k(x - c(t)),
\end{equation}
where $\lambda(t)$ is now playing the role of the center of the harmonic
potential $c(t)$. Initially, we consider the case with a single learnable parameter
$N_p=1$, denoted as $c_m$. This scenario resembles a two-step protocol for
stiffness, but is now applied to the center. Let $c_i$ and $c_f$, the initial
and final values of the center, respectively. Then, the two-step protocol for
the center is
\begin{equation}
  c(t) = 
  \begin{cases}
    c_i, & \text{if }  t \leq 0, \\
    c_m, & \text{if } 0 < t < t_f, \\
    c_f, & \text{if } t \geq t_f.
  \end{cases}
\end{equation}
For the force Eq.~\eqref{eq:force center} and initial condition $P(x,0)=e^{-k(x-c_i)^2/2}\sqrt{k/(2\pi)}$, the Fokker-Planck equation admits a solution
\begin{equation}
P(x, t) = \frac{1}{\sqrt{2\pi/k}} e^{-\frac{k}{2}\left( x- \langle x(t) \rangle\right)^2}
\end{equation}
with
\begin{equation}
  \langle x(t) \rangle = c_m + (c_i-c_m)e^{-kt} 
\end{equation}
for $0 < t < t_f$. The adjustable parameter $c_m$ is determined assuming
equilibrium at $t=t_f$, that is, the average position,  $\langle x(t) \rangle$, is $c_f$ at $t_f$. This
gives rise to the value of
\begin{equation}\label{eq:condition_c_m}
c_m = \frac{c_f-c_i e^{-k t_f}}{1 - e^{-k t_f}}.
\end{equation}
If the duration of the protocol $t_f$ is shorter than the natural relaxation time, we achieve a fast thermal equilibration.

\begin{figure}[ht]
  \centering
  \includegraphics[width=0.99\textwidth]{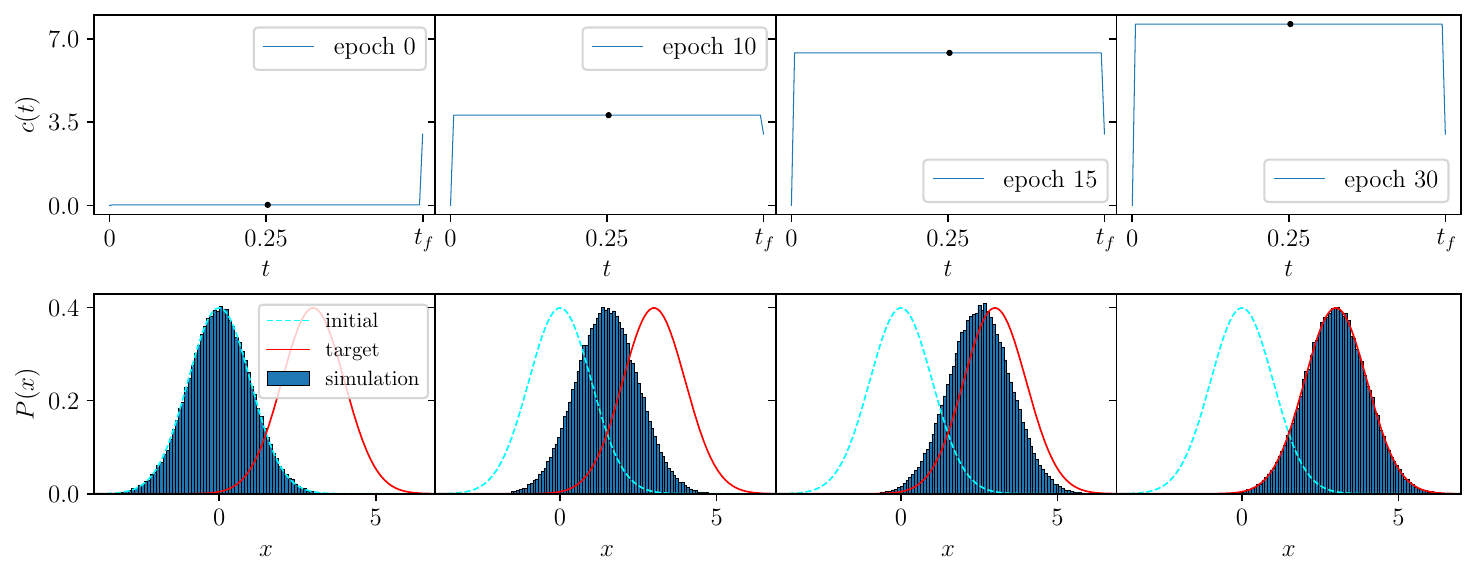}

  \caption{Training of a single-parameter model to replicate the two-step
  protocol for the center for fast thermal equilibration, with $c_i=0.0$,
  $c_f=3.0$ and $t_f=0.5$. The initial and target probability distributions are displayed, alongside a histogram generated from the model trained up to the despicted epoch.
  \label{fig:change-center}}
  
\end{figure}

Using a piecewise linear model with a single adjustable parameter $ N_p = 1$ and
training the model using the loss function defined in Eq.~\eqref{eq:loss-char},
we can compare the value of the adjustable parameter to the solution of
Eq.~\eqref{eq:condition_c_m} to evaluate the predictive accuracy of the
algorithm. The training process was carried out with the following parameters:
$c_i=0$, $c_f=3.0$, $t_f=0.5$ for 30 epochs with a learning rate of $\eta=10$.
For this set of parameters, the characteristic relaxation time $\tau=1/k=1.0$.
We choose a protocol duration $t_f=0.5$ smaller than the relaxation time. The
results of training are depicted in Fig.~\ref{fig:change-center}. This figure
shows the initial and final probability density functions after each epoch.
After time \( t_f \), the system reaches the required equilibrium state. During
training, the loss function was approximately $10^{-2}$ at epoch 15, and the
training continued until epoch 30, when the loss function decreased to
approximately $10^{-5}$. The trained value of \( c_m \) obtained through this
process was \( c_m = 7.6241 \), compared to the expected value of \( c_m =
7.6245 \). This results in a deviation of approximately \(0.004\%\) from the
expected value.

\begin{figure}[ht]
  \centering
  \includegraphics[width=0.99\textwidth]{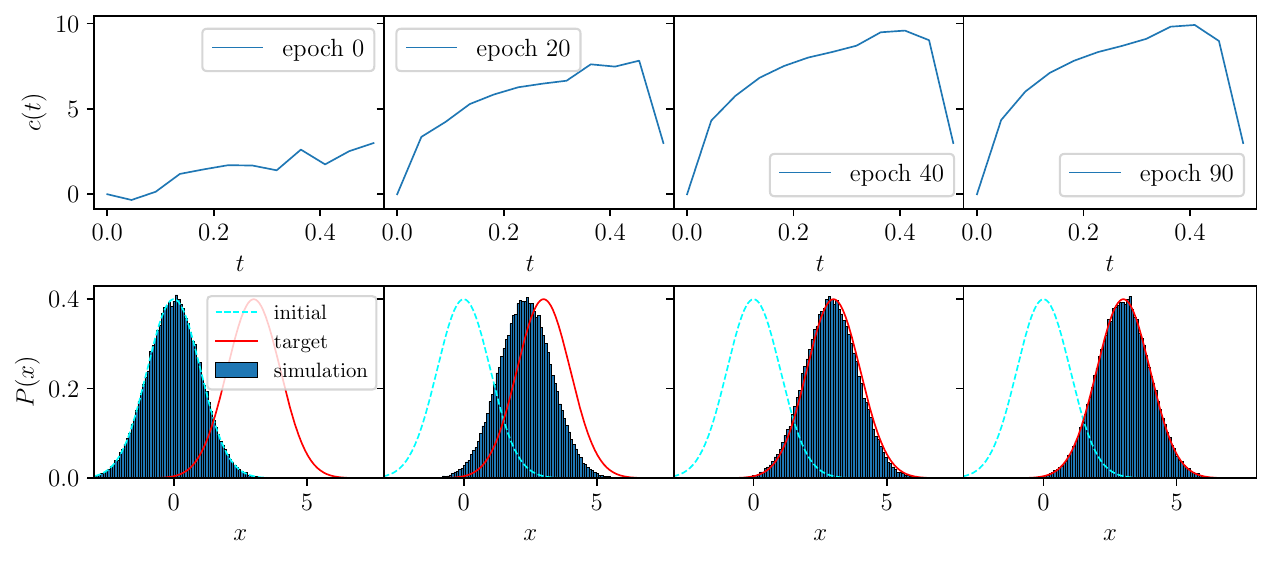}
  \caption{Training for the center for fast thermal equilibration using a $10$ parameter model and the loss function include the penalty term $L_{gradc}$. The initial (solid), target(dashed), and histogram of the trained model are displayed after some epochs. The parameters were $c_i=0$, $c_f=3.0$ and $t_f=0.5$.
   \label{fig:center-N30}}

\end{figure}

Now, we consider the general case where the center is an arbitrary function of
time. In this situation, no analytical solution exists in general. The center is
parameterized using a piecewise linear function with $N_p = 10$ learnable
parameters. This model assumes that $c(t)$ is a continuous function in $t=0$ and
$t=t_f$. The model is trained with a loss function that is a convex linear
combination of $L_{char}$ and a penalty term $L_{gradc}$. The term $L_{gradc}$
penalizes strong variations in $c(t)$ and has the same structure as
Eq.~\eqref{eq:loss_k_varitaions} for $k(t)$. The model underwent two training
stages. The first stage, which lasted for 50 epochs, used a blend parameter of
$\alpha=10^{-3}$. The second stage, which lasted for 40 epochs, used $\alpha=0$.
Both stages used a learning rate of $\eta=50$. The training process for selected
epochs is illustrated in Fig.~\ref{fig:center-N30}, which displays the
probability density functions at various epochs. At epoch 50, the loss function
was about $10^{-3}$, and training continued until epoch 90, when the loss
function reached the order of $10^{-6}$, indicating that the system is in the
final equilibrium state. As shown in Fig.~\ref{fig:center-N30}, the model
achieves high precision and the probability density function at the end aligns
with the expected equilibrium distribution.

\begin{figure}[ht]
  \centering 
  \includegraphics[width=0.5\textwidth]{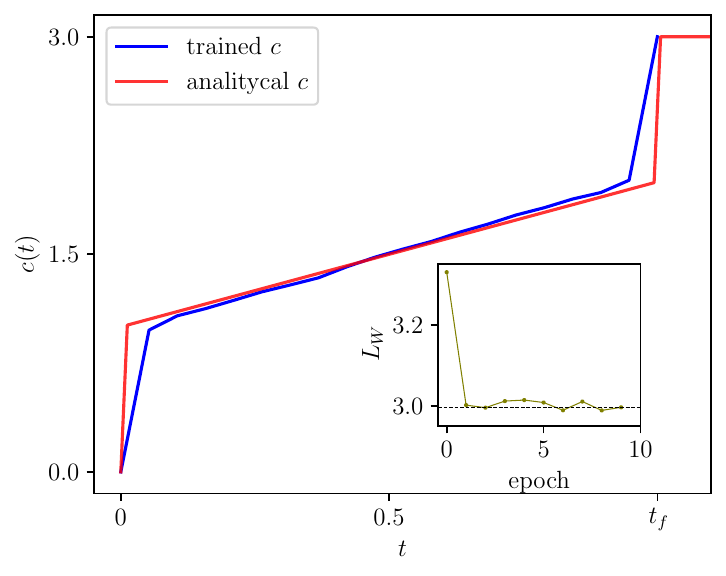}
  \caption{Center position as a function of time to optimize the work done on the
  system. The red curve is the analytical solution
  Eq.~\eqref{eq:optimal_protocol_and_work_center}, and the blue line is the
  trained protocol. The inset illustrates the loss function
  Eq.~\eqref{eq:average_work} during the training, while the dotted line
  indicates the optimal work, as in
  Eq.~\eqref{eq:optimal_protocol_and_work_center}. The initial and final
  positions were $c_i = 0.0$ and $c_f = 3.0$, respectively. The final time was
  fixed at $t_f = 1.0$, and the stiffness was set to $k =
  1.0$.\label{fig:center_protocol_opt_work}} 
\end{figure}

We now examine the scenario where the loss function is characterized by the average work, as defined in Eq.~\eqref{eq:average_work}. As demonstrated in Ref.~\cite{Seifer-optimal}, the optimal protocol $c^{*}_0(t)$ and the optimal work $\expval{W}_{opt}$ are
\begin{equation}\label{eq:optimal_protocol_and_work_center}
    c^{*}_0(t) = \frac{c_f(t+1)}{t_f+2}, \quad
    \expval{W}_{\text{opt}} = \frac{c_f^2}{t_f+2} 
\end{equation}
for $0<t<t_f$. It is evident that the protocol $c^{*}_0(t)$ exhibits discontinuities in $t = 0.0$ and $t = t_f$. 
As indicated in Fig.~\ref{fig:center_protocol_opt_work}, the precision in determining the optimal protocol is notably high in just 10 epochs and a learning rate of $\eta=10$. For the set of parameters $c_i=0$, $c_f=3.0$ and $t_f=1.0$, the theoretical value of the optimal work given by equation~\eqref{eq:optimal_protocol_and_work_center} is $\langle W \rangle_{\text{opt}} = 3.0$, while the result obtained through training after 10 epochs and a learning rate of $\eta=10$ was $\expval{W}= 3.0060$. These results imply a relative error of $0.2\%$.

\subsection{The center and stiffness of the harmonic potential as a control parameter}
In this section, we consider the simultaneous adjustment of both the center and stiffness, which influence the system's dynamics through a non-linear coupling described by the force in Eq.~\eqref{eq:force center}. As an initial example, we analyze a TSP for both parameters, where the stiffness $k_m$ is determined by Eq.~\eqref{eq:TSP-k_m} and the center $c_m$ by Eq.~\eqref{eq:condition_c_m} with $k = k_m$.

The stiffness and center are modeled using a piecewise linear parameterization with $N_p = 1$ for each. Given that these parameters contribute differently to the dynamics—where $k$ multiplies $x$ while $c(t)$ does not—their gradients affect the optimization process differently. Consequently, different learning rates are required for each parameter.

For numerical simulations, we set $c_i = 1.0$, $k_i = 0.5$, $c_f = -1.0$, and
$k_f = 1.0$. Based on these parameters, the simulations suggest a relaxation
time of order $\tau = 4.0$. We aim to accelerate thermal equilibration in a
final time $t_f = 0.1$. The theoretical values for $k_m$ and $c_m$, obtained as
solutions to Eqs.~\eqref{eq:TSP-k_m} and~\eqref{eq:condition_c_m}, are $k_m =
4.1931$ and $c_m = -4.8394$.

\begin{figure}
    \centering
    \includegraphics[width=0.95\textwidth]{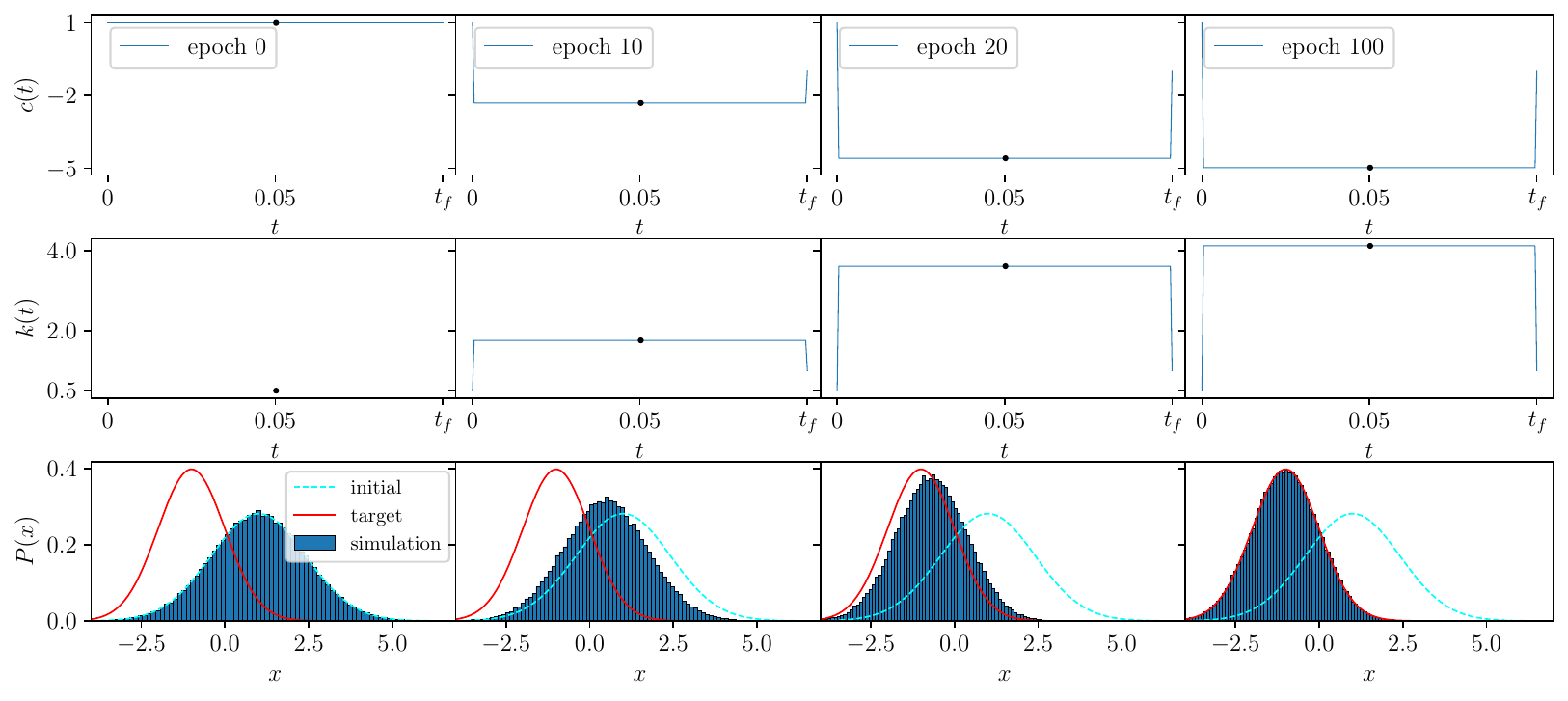}
    \caption{ Evolution of training of a model to reproduce the TSP for stiffness and center simultaneously. The initial (solid) and target (dashed) distributions correspond to equilibrium states, each described by Gaussian distributions with center $c$ and variance $1/\sqrt{k}$. The histogram was generated with the trained model until the epoch shown. The parameters were $c_i =-c_f= 1.0$, $k_f = 1=2k_i$, and $t_f=0.1$.}
    \label{fig:TSP_k_and_c}
\end{figure}

The model was trained for 100 epochs using learning rates $\eta_c = 10.0$ for the center and $\eta_k = (2/3) \eta_c$ for stiffness, resulting in a TSP solution with $k_m = 4.1218$ and $c_m = -4.9733$. The corresponding relative errors are $1.69\%$ and $2.76\%$, respectively. The results presented in Fig.~\ref{fig:TSP_k_and_c}, confirm that the equilibrium state is successfully reached at $t_f$ by the final epoch.  

The more general case is where the stiffness and center are not constant in $[0, t_f]$. To address this scenario, we parameterize each variable using a piecewise linear model with $N_p = 10$, ensuring that the control variables remain continuous in $[0, t_f]$. Given that strong variations are expected at the beginning and end of the protocol, we introduce a penalty term in the loss function, $L_{\text{grad}k} + L_{\text{grad}c}$, to mitigate abrupt changes in $k$ and $c$ over the time interval $[0, t_f]$.

\begin{figure}
    \centering
    \includegraphics[width=0.99\linewidth]{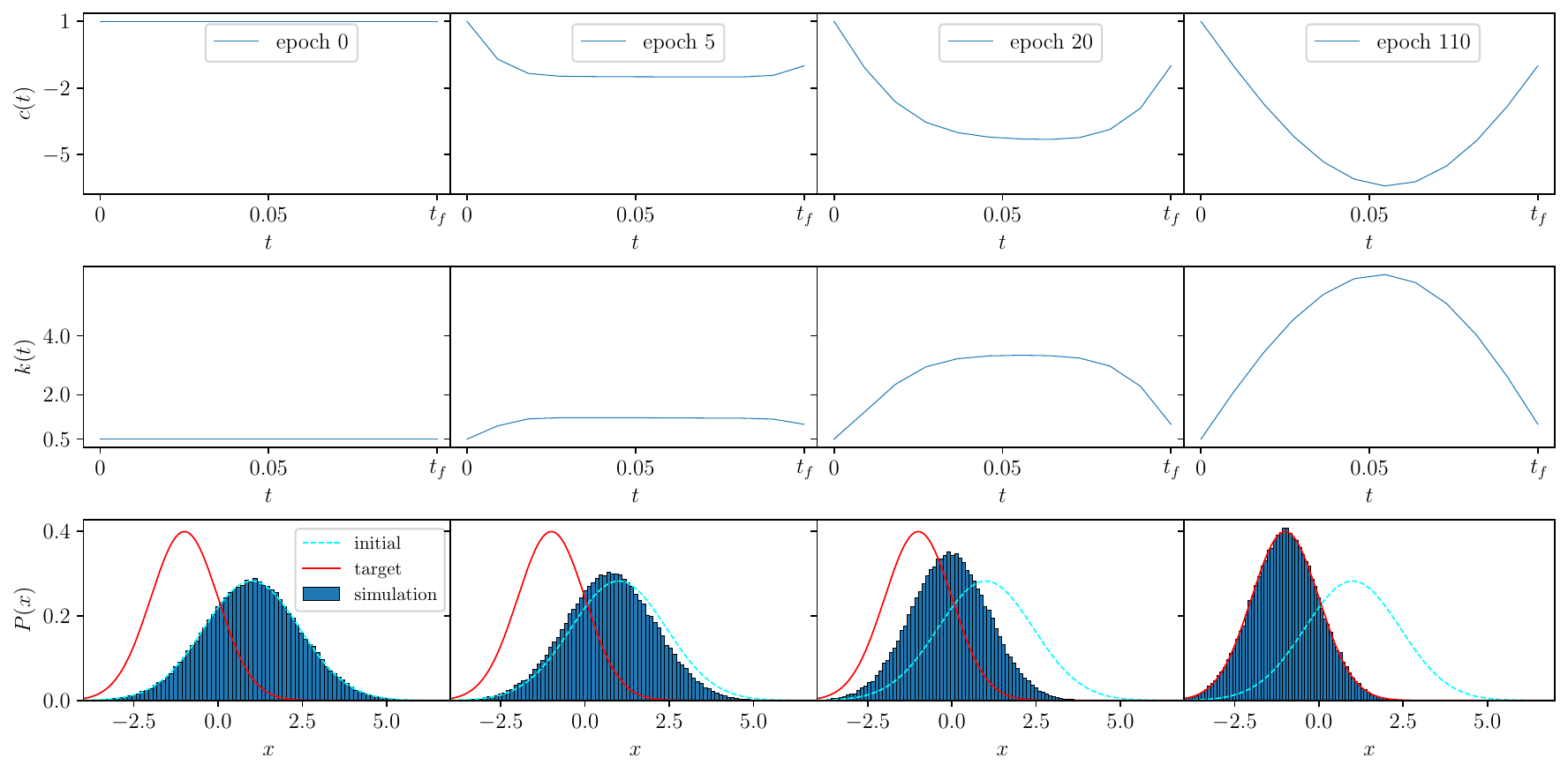}
    \caption{The training evolution of a fast thermal equilibration model
    incorporating 10 parameters for stiffness and 10 parameters for center is
    shown. The loss function employed is a convex linear combination of
    $L_{char}$ and $L_{\text{grad}k} + L_{\text{grad}c}$. The system is
    initially trained for 100 epochs using a blend parameter of $10^{-2}$,
    followed by an additional 10 epochs with the loss function $L_{char}$ to
    emphasize the fast thermal process. The parameter values were set as $k_f =
    2k_i = 1.0$ and $c_f = -c_i = -1.0$, with a learning rate of $\eta_c = (3/2)
    \eta_k = 100$. }
    \label{fig:TSP_k_and_c_control_vars}
\end{figure}

The model was initially trained for 100 epochs using the total loss function with a blend of $\alpha=10^{-2}$, followed by an additional 10 epochs trained exclusively with $L_{char}$. Representative epochs are depicted in Fig.~\ref{fig:TSP_k_and_c_control_vars}, which illustrates that both $k$ and $c$ exhibit smooth variations and continuity. The final epoch confirms that the system successfully reaches equilibrium at $t = t_f$.

Through these examples, we have demonstrated that the algorithm successfully
finds optimal protocols within the harmonic framework, effectively reproducing
established analytical results. Furthermore, its applicability extends to the
discovery of more complex protocols, as explored in the following section.

\section{Brownian particle under nonharmonic potentials}
\label{sec:brownian-nongauss}

With minimal modifications, our piecewise linear parameterization model can be
adapted to tackle more complex problems where analytical methods provide limited
or no insight. As discussed in Ref.~\cite{Seifer-optimal}, solving optimization
problems for non-harmonic potentials presents significant challenges, often
rendering analytical solutions impractical and necessitating numerical
techniques from optimal transport theory~\cite{Optimal_Protocols_Optimal_Transport,Baldovin2025}.

\begin{figure}[ht]
  \centering 
  \includegraphics[width=0.7\textwidth]{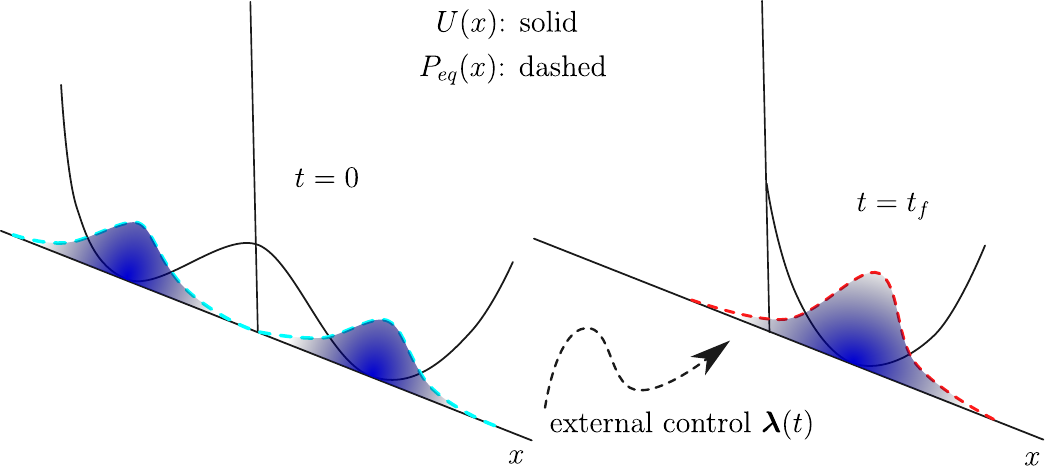}  \caption{Illustration of the externally controlled deformation of a fourth-degree potential at $t=0$ into a parabolic potential at $t=t_f$. Multiple pathways $\boldsymbol{\lambda}(t)$  exist to achieve this transformation.\label{fig:double-well}}
\end{figure}

In this section, we demonstrate how our algorithm extends beyond the harmonic
potential to address fast-thermalization problems. Specifically, we investigate
the design of fast thermal protocols for transforming a fourth-degree polynomial
potential energy at $t=0$ into a quadratic potential at $t_f$. Given the
importance of double-well potentials and their deformation into quadratic
potentials in the context of memory erasure~\cite{info-Bechhoefer, info-termo},
we consider an initial potential with a double-well shape. The final state can
be any confining potential; as an illustrative example, we select a quadratic
potential obtained from the Taylor expansion of the initial potential around one
of its minima (e.g., the right minimum). However, alternative final states can
also be chosen. See Fig.~\ref{fig:double-well} for details.
Similar to the harmonic case, there exist infinitely many possible paths
connecting the initial and final states. Thus, our goal is to identify an
external protocol that accelerates equilibration relative to the natural
relaxation time.

We consider a time-dependent potential energy of the form:
\begin{align}\label{eq:four degree potential}
U(x, \lambda(t)) = a_4(t) x^4 + a_3(t) x^3 + a_2(t) x^2 + a_1(t) x + a_0(t),
\end{align}
where the coefficients $\lambda=\{a_0,a_1,a_2,a_3,a_4\}$ are functions of time and serve as control parameters.
A straightforward approach would be to parameterize each coefficient and train the model to minimize $L_{char}$. However, this method presents challenges. During training, the parameters are updated following Eq.~\eqref{eq:gradient-descent-general}, which can result in a repulsive potential energy, causing the particle to escape to infinity. This issue not only affects numerical simulations but also poses a fundamental obstacle to the experimental realization of the protocol. This can be understood in the following way. The force act during the training is
\begin{equation}
    F(x,\lambda(t))=-(4a_{4}(t)x^3 + 3a_{3}(t)x^2 + 2a_2(t)x +a_{1}(t)).
\end{equation}
If the algorithm updates the parameter $a_{4}$ to be negative for some time (or
times), the force tends to move the particle to the $|x|\rightarrow\infty$ directions,
then the algorithm diverge. If the parameter $a_4$ is very close to zero by the
positive numbers and the coefficient $a_3>0$, the coefficient $a_3$ makes the
main contribution to the force. Therefore, if the particle is located in $x<0$,
it has a tendency to move to the left, that is, it moves away from the expected
final position distribution, which in this case is around $x=1.0$. This behavior
leads to instabilities and convergence problems. Thus, we impose an additional
constraint, given by $a_3(t)<0$, a constraint motivated purely by the numerical
convergence. In summary, our model included $a_{4}(t)>0$ and $a_{3}(t)<0$ in $t
\in (0,t_f)$ as restrictions.
To address this, the model must incorporate constraints that ensure that the potential remains confining and converge. This can be achieved either by explicitly enforcing confinement conditions within the model or by introducing a penalty term in the loss function to discourage nonphysical solutions. We used a model with the restrictions incorporated.

To find the fast-thermalization protocol while ensuring confinement, we trained
a model with two different loss functions. In both approaches, we modeled each
coefficient with a piecewise linear parametrization with $N_p=20$ parameters and
including the restriction $a_{4}(t)>0$ and $a_{3}(t)<0$ in $(0,t_f)$. In
addition, we chose the potential at $t=0$ to have two symmetric minima at
$x=-1.0$ and $x=1.0$ with energy $0$, a maximum at $x=0.0$ with a separation
barrier $U(0,0)=4.0$. At the end of the protocol, we selected the potential at
$t=t_f$ to be a harmonic potential with minimum centered at $x=1.0$
corresponding to the Taylor expansion to second order of the initial quartic
potential around $x = 1.0$. To determine the system's relaxation time, we
proceed as in the harmonic case, that is, the initial potential is abruptly
switched to the final one (step protocol), and the time required for the system
to reach equilibrium is observed. For this set states, the relaxation time was
$\tau = 0.6$. Hence, we established the value of $t_f$ as $0.06$, making the
protocol approximately 10 times faster than the relaxation time. 

It is important to note that in this section we employed the Adam optimizer rather than stochastic gradient descent (SGD). This choice was motivated by the observation that the coefficients in the quartic potential play different roles in ensuring the boundedness of the potential. Consequently, it is essential to exercise precise control over the manner in which these parameters are updated during the training process. Adam proved to be more robust in this context because it utilizes adaptive learning rates and incorporates momentum, allowing for individualized parameter updates \cite{Adam_optimizer}. This adaptive mechanism is particularly beneficial when dealing with parameters that vary in importance, as it facilitates efficient convergence and helps mitigate the risk of stagnation during training.

\begin{figure}
  \includegraphics[width=0.32\textwidth]{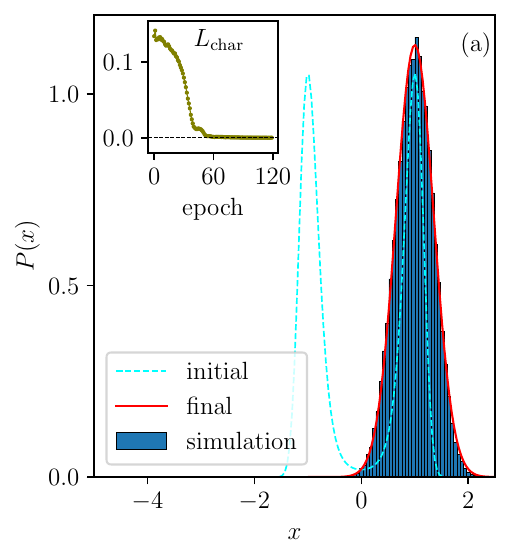}
  \includegraphics[width=0.33\textwidth]{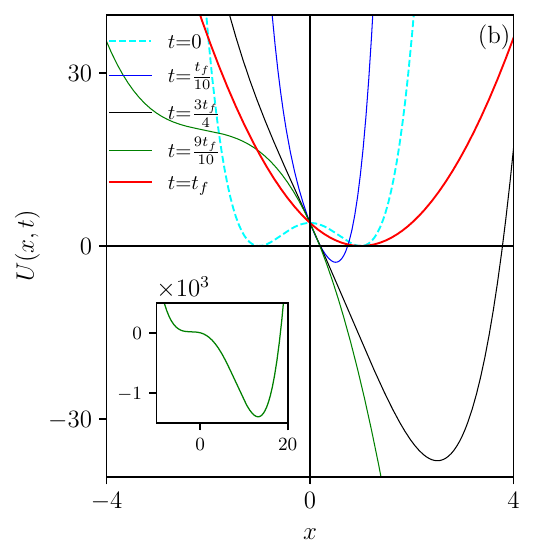}
  \includegraphics[width=0.33\textwidth]{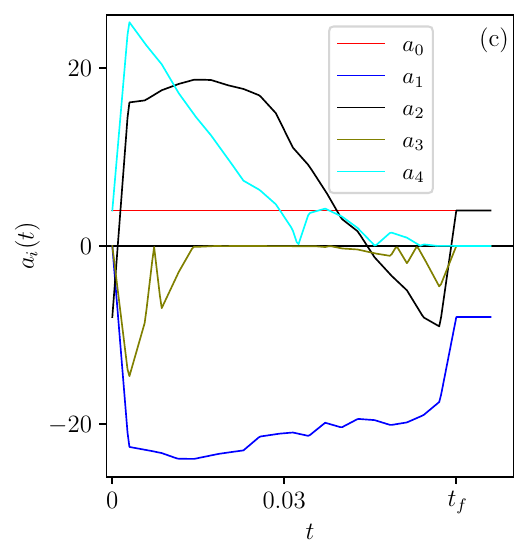}
  \caption{Fast thermal protocol that transforms a quartic potential into a
  harmonic potential without smoothing the parameters changes.    
  (a) Histogram of the trained model in the last epoch, i.e., epoch 120. The dotted line represents the initial quartic potential, while the solid line corresponds to the final harmonic potential. The inset shows the loss function as a function of epochs. (b) Some snapshots of the potential energy of the trained model. (c) Coefficients of the trained model as a function time.
\label{fig:quartic-fastthermal-histo}}
\end{figure}

This model was trained for 120 epochs with the loss function Eq.~\eqref{eq:loss-char} and a learning rate of $\eta=0.5$. 
The histogram of the trained model is shown in Fig.~\ref{fig:quartic-fastthermal-histo}~(a). At the end of the training process, the loss function reaches a value on the order of $10^{-4}$ (inset plot of Fig.~\ref{fig:quartic-fastthermal-histo}~(a)), which means that the system has reached equilibrium. As illustrated in Fig.~\ref{fig:quartic-fastthermal-histo}~(b), the potential evolves to significantly higher values compared to its initial state, as highlighted in the inset. This increase in $\abs{U}$ reflects the high probability that the particle will be found in these regions, with the external potential effectively guiding it toward the final equilibrium configuration. Despite the fact that this protocol is fast thermal, the figure~\ref{fig:quartic-fastthermal-histo}~(c) shows abrupt changes in the coefficients, a feature that is not desirable from the experimental point of view.

This behavior mirrors that observed in the harmonic case. Following the same ideas as those for the harmonic case, we introduced a penalty term in the loss function to discourage large variations in the coefficients. 
The training process was divided into two stages. In the first stage (up to epoch 100), we employed the loss function:
\begin{align}
L &= (1-\alpha)L_{char} + \alpha L_{grad\,\text{a}},\\
L_{grad\,\text{a}} &=\sum_{i=0}^{4} \sum_{n=0}^{N_p-1}
\left(a_{i}(t_{n+1}) - a_{i}(t_n) \right)^2
\end{align}
where the blend parameter was set to $\alpha = 10^{-3}$ and the learning rate to $\eta = 0.5$. At the end of this stage, the loss function was of order $10^{-3}$.

\begin{figure}
  \includegraphics[width=0.32\textwidth]{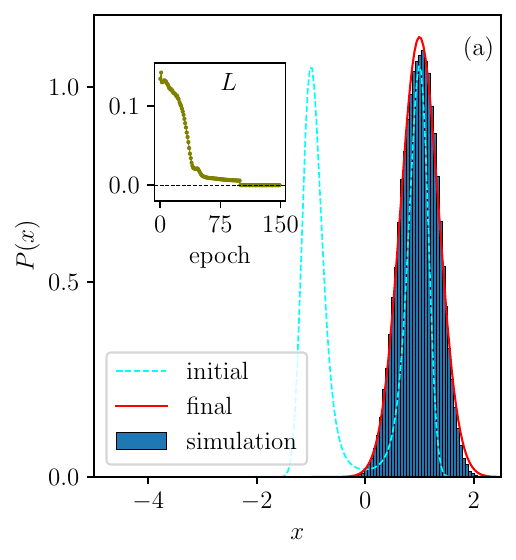}
  \includegraphics[width=0.32\textwidth]{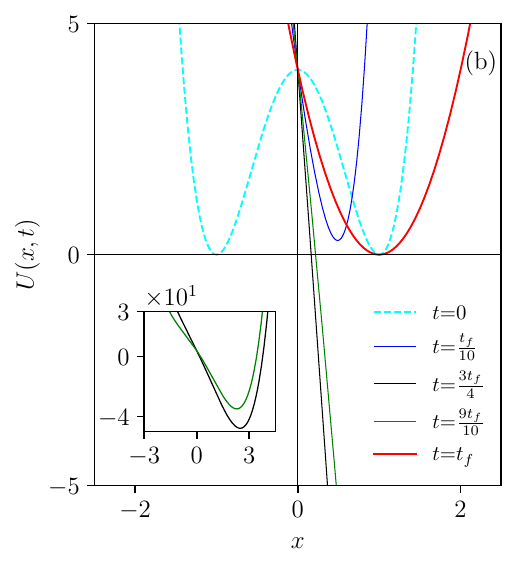}
  \includegraphics[width=0.32\textwidth]{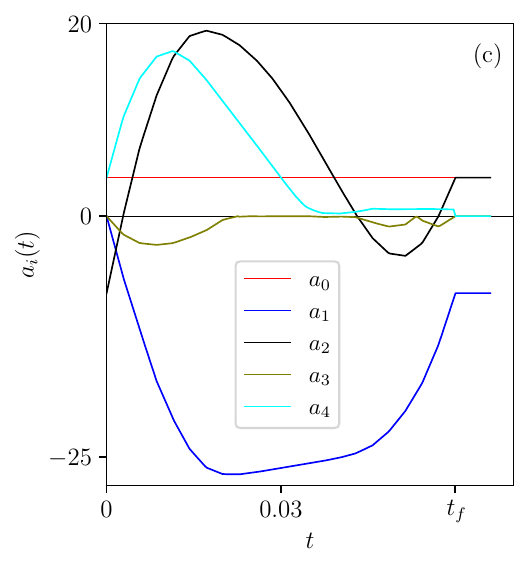}
  \caption{Fast thermal protocol that transforms a quartic potential into a
  harmonic potential with smoothing the parameters changes.   (a) Histogram of the trained model in the last epoch, i.e., epoch 120. The dotted line represents the initial quartic potential, while the solid line corresponds to the final harmonic potential. The inset shows the loss function as a function of epochs. (b) Some snapshots of the potential energy of the trained model. (c) Coefficients of the trained model as a function time.
\label{fig:quartic-fastthermal-histo-control-vars}}
\end{figure}
In the second stage, the model was further trained for 50 epochs using only the loss function $L_{char}$ (i.e., setting $\alpha = 0$) to focus entirely on optimizing fast-thermal equilibration, the final value of the loss function was $10^{-5}$.
 
Figure~\ref{fig:quartic-fastthermal-histo-control-vars} resumes the results for this case. Figure~\ref{fig:quartic-fastthermal-histo-control-vars}~(a) presents the histogram obtained with the trained model, and the inset describes the evolution of the loss function. In figure~\ref{fig:quartic-fastthermal-histo-control-vars}~(b) selected snapshots of the optimized potential were plotted, but remarkably, the variations in $U$ are two orders of magnitude less than in the case without the penalty term (see the inset of Fig.~\ref{fig:quartic-fastthermal-histo-control-vars} (c)). In the figure~\ref{fig:quartic-fastthermal-histo-control-vars}~(c) is plotted the temporal evolution of the trained potential coefficients, demonstrating the expected smoothing effect introduced by the penalty term. The gradual variation of the coefficients ensures a more stable and controlled evolution of the potential, which is particularly advantageous for experimental implementations, where abrupt parameter changes can be challenging to realize.
 
Both implementations successfully achieve the desired transformation, although the penalty-based approach offers a more feasible solution. These findings highlight the flexibility of our framework in adapting to different physical constraints, paving the way for practical implementations in experimental settings.




\section{Conclusions}

In this study, we demonstrated how to apply machine learning tools to find control protocols that optimize conditions such as mean work or protocols that achieve equilibrium faster than the relaxation time, or both conditions taken simultaneously. From the practical point of view, we have shown the flexibility that this numerical approach offers. It is possible to parametrize protocols in different ways or include restrictions in the models that achieve a specific goal. These restrictions cannot be arbitrary but motivated from physical reasons. Otherwise, a machine learning algorithm could find solutions that lack physical meaning. 
 
We reproduced the well-established results for the stiffness of the harmonic oscillator in the overdamped limit. These results have been reproduced using different model parametrizations to recall the flexibility of our approach. In addition, we showed that our method is also suitable for dealing with the center of the harmonic potential, indicating independence of the parameters used to perform the control on the stochastic system. Furthermore, we studied non-harmonic potential, and we concluded that our approach is perfectly well adapted to more complex potentials. This is one of the most important results of this work. The scalability and adaptability of machine learning for stochastic systems offer a new numerical tool to deal with the complexities of stochastic control theory and stochastic thermodynamics. 
 
While our approach offers several advantages, it also has certain limitations. Perhaps the most significant challenge is extending the framework to systems with multiple interacting particles. This requires solving coupled stochastic differential equations, often with different external control parameters, where analytical methods are virtually nonexistent. In addition, our approach is currently limited to the overdamped regime, and the inclusion of inertia represents some challenges~\cite{gomez-marinOptimalProtocolsMinimal2008,muratore-ginanneschiHowNanomechanicalSystems2014,barrosLearningEfficientErasure2025,Baldovin2025}. Developing generalized algorithms capable of handling interactions and the underdamped regime remains an important direction for future work.
 
Another limitation is the lack of direct physical interpretation for some parameters, particularly those emerging from the neural network approach. The internal parameters of the network do not always correspond to physically meaningful quantities, which can make the results less intuitive. Additionally, there is considerable flexibility in selecting functions such as $L_{char}$ and the penalty terms, which introduces an element of arbitrariness.
 
Despite these challenges, our results provide a foundation for exploring how machine learning techniques can be effectively applied to stochastic control dynamics and thermodynamics. Future work in this direction could lead to more refined and physically interpretable models, further bridging the gap between machine learning and statistical physics.

The code we developed in this work is open source and it is available at~\cite{langesim-optim} as a \texttt{python} library which can be used to reproduce our results, train new protocols and extend it for more complex potentials. The documentation of the code can be found at~\cite{langesim-optim-doc}.

\section{Acknowledgments}
We gratefully acknowledge the support of Universidad de los Andes, Facultad de
Ciencias, projects number INV-2025-205-3271 and INV-2023-176-2951, and its High
Performance Computing center (HPC). We acknowledge support from ECOS Nord
project C24P01 and Minciencias, Patrimonio Autónomo Fondo Nacional de
Financiamiento para la Ciencia, la Tecnología y la Innovación, Francisco José de
Caldas. We thank the Deutscher Akademischer Austauschdienst (DAAD) for financial support. We also thank Sabine Klapp and her research group for inspiring discussions and hospitality at Technical University Berlin, Germany.

\clearpage 
\bibliographystyle{apsrev4-2}

\bibliography{biblio}

\end{document}